\documentclass{article} % For LaTeX2e
\usepackage[preprint]{colm2026_conference}

\usepackage{microtype}
\usepackage{hyperref}
\usepackage{url}
\usepackage{booktabs}
\usepackage{times}
\usepackage{soul}
\usepackage{url}
\usepackage[T1]{fontenc}
\usepackage[utf8]{inputenc}
\usepackage[small]{caption}
\usepackage{graphicx}
\usepackage{amsmath}
\usepackage{amsthm}
\usepackage{booktabs}
\usepackage{algorithm}
\usepackage{algorithmic}
\usepackage[switch]{lineno}
\usepackage{natbib}
\usepackage{amssymb} 
\usepackage{enumitem}
\usepackage{subcaption}
\usepackage{graphicx}
\usepackage{wrapfig}
\usepackage{listings}
\usepackage{enumitem}
\usepackage[most]{tcolorbox} %     
\usepackage{graphicx}
\usepackage{makecell}

\usepackage{lineno}

\definecolor{darkblue}{rgb}{0, 0, 0.5}
\hypersetup{colorlinks=true, citecolor=darkblue, linkcolor=darkblue, urlcolor=darkblue}

\title{SimCity: Multi-Agent Urban Development Simulation \\ with Rich Interactions}

\author{
Yeqi Feng$^{*1,2}$,
Yucheng Lu$^{*4}$,
Hongyu Su$^{1,2}$,
Yixin Tao$^{5}$\&
Tianxing He$^{\dagger 1,2,3}$\thanks{Equal contribution. $^\dagger$ Corresponding authors.}\\   
$^1$Institute for Interdisciplinary Information Sciences, Tsinghua University\\
$^2$Shanghai Qi Zhi Institute\\
$^3$Xiongan AI Institute\\
$^4$New York University\\
$^5$Shanghai University of Finance and Economics\\
\texttt{\{fengyq25, suhy25\}@mails.tsinghua.edu.cn}
 \\
\texttt{yl6586@nyu.edu} \\
\texttt{hetianxing@mail.tsinghua.edu.cn} \\
}

\begin{document}

\ifcolmsubmission
\linenumbers
\fi

\maketitle

\begin{abstract}
Large Language Models (LLMs) open new possibilities for constructing realistic and interpretable macroeconomic simulations. We present \textbf{SimCity}, a multi-agent framework that leverages LLMs to model an interpretable macroeconomic system with heterogeneous agents and rich interactions. Unlike classical equilibrium models that limit heterogeneity for tractability, or traditional agent-based models (ABMs) that rely on hand-crafted decision rules, SimCity enables flexible, adaptive behavior with transparent natural-language reasoning. Within SimCity, four core agent types (households, firms, a central bank, and a government) deliberate and participate 
in a frictional labor market, a heterogeneous goods market, and a financial market. Furthermore, a Vision–Language Model (VLM) determines the geographic placement of new firms and renders a virtual city map, allowing us to study both macroeconomic regularities and urban expansion dynamics within a unified environment. To evaluate the framework, we compile a checklist of canonical macroeconomic phenomena, including price elasticity of demand, Engel’s Law, Okun’s Law, the Phillips Curve, etc., and show that SimCity effectively and robustly exhibits these empirical patterns. Using our framework, we study novel economic shocks that are difficult to analyze in existing models and examine their macroeconomic implications, yielding economically coherent results.

\end{abstract}

\section{Introduction}

% 第三段：用大语言模型进行多智能体的模拟工作（包括博弈论等）
% 第一段：AI 经常用于传统经济学研究，ABM 是宏观经济学研究的重要方法
% 第四段：EconAgent 探讨了用大语言模型进行宏观经济学模拟的可能性，但它的模型设计较为简陋。
% 第五段：为了模拟更多的现象，我们进行了若干建模和实验设计。工作的贡献如下
% 1. 对先前工作进行了补充，We design the first LLMs driven hetergenous ABM. Containing firm and its behaviors
% 2. We list a checklist and evaluate traditional Economic Simulation and LLMs based ABM
% 3. We conduct macroeconomic simulations and run 一系列 experiment 说明 our framework provides a new approach to 经济学研究。

%Artificial Intelligence (AI), particularly
The rapid development of Large Language Models (LLMs) has enabled multi-agent simulations of human societal activities across diverse scales and domains \citep{Gao2024LLMAgentSurvey}. In these settings, autonomous LLM-powered agents interact with each other and with their environment. Existing work has examined not only general social simulacra  \citep{park2023generativeagentsinteractivesimulacra, huang2025adasocietyadaptiveenvironmentsocial, piao2025agentsocietylargescalesimulationllmdriven}, but also domain-specific applications such as public administration crisis \citep{xiao2023simulatingpublicadministrationcrisis}, health policy~\citep{hou2025societygenerativeagentssimulate}, or deduction game~\citep{xu2025languageagentsreinforcementlearning}. While these works demonstrate the social simulation capabilities of LLM-driven agents, we focus on their application to the simulation and evaluation of urban-style economic activities.

For the past two decades, dynamic stochastic general equilibrium (DSGE) models have been the predominant paradigm for studying aggregate economic behavior \citep{Blanchard2009StateOfMacro, Glandon2023MacroeconomicResearch}. While mathematically elegant, DSGE relies on the assumption of simplified representative agents solving explicit optimization problems, which restricts heterogeneity and richer behavioral dynamics \citep{Sergi2018DSGE, Vines2020RebuildingMacroII, Storm2021CordonConformity}. Agent-based models (ABMs) offer a bottom-up alternative that accommodates heterogeneity but typically rely on hand-crafted decision or learning rules \citep{Dilaver2018AgentBased}.
Advancing the ABM tradition with recent developments in LLMs, we propose \textbf{SimCity}, a multi-agent macroeconomic simulation framework that enables flexible, adaptive behavior with transparent natural language reasoning, and a virtual city environment.

\begin{figure*}[t]
\centering
\includegraphics[width=\textwidth]{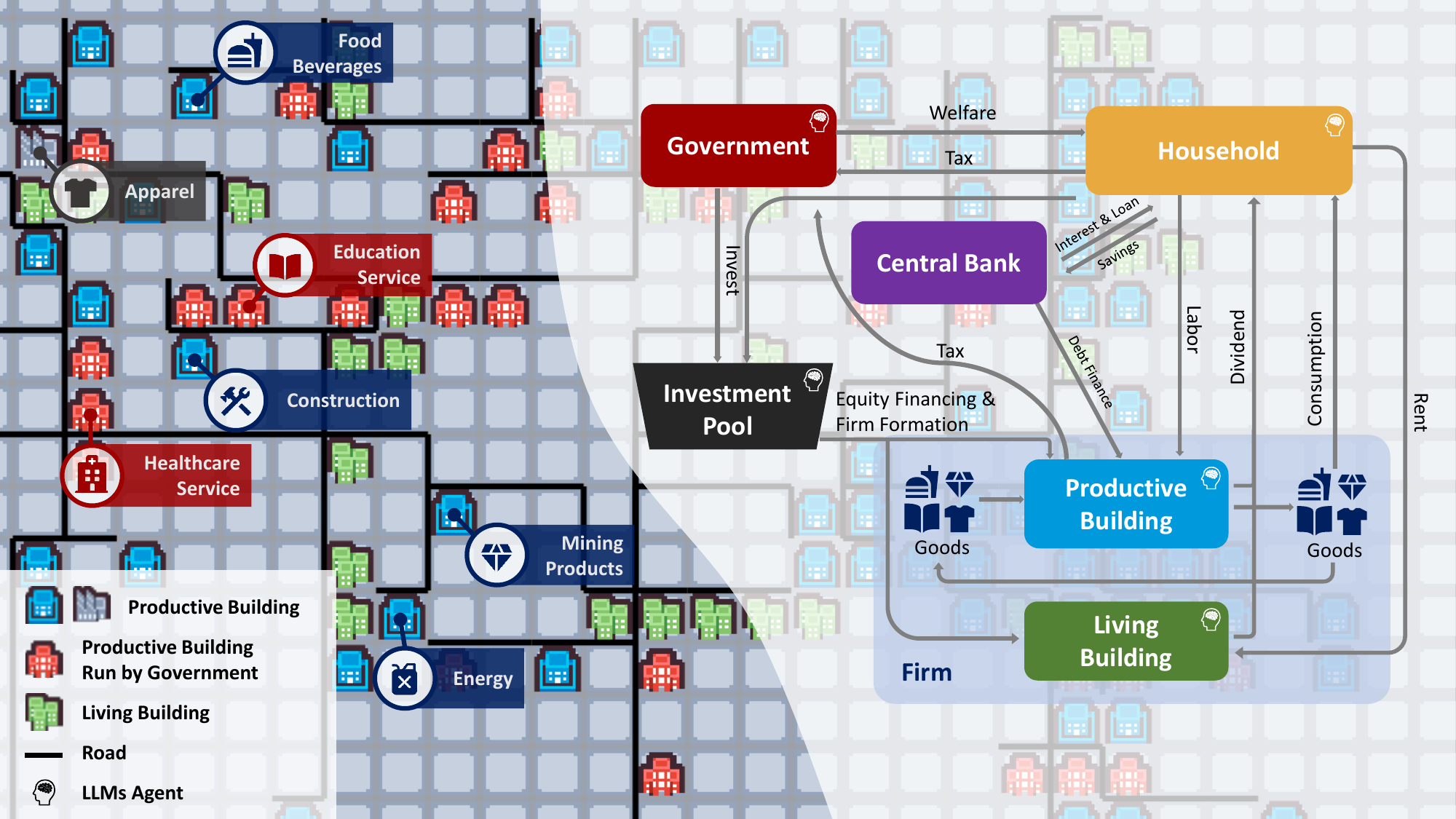}
\caption{The SimCity framework. Left: A visualized map illustrating the urban layout across three building categories. Households reside in living buildings and engage in labor within productive buildings. City evolution is depicted in Figure~\ref{fig::phase1}. Right: Diagram of multi-agent interactions. Gray arrows denote the flow of capital and goods. Mechanisms of interactions are detailed in Section~\ref{sec::modeling}.}
\label{fig::pipeline}
\end{figure*}

% To split
SimCity models the economy as the interaction of four types of agents: households, firms, a central bank, and a government. To incorporate realistic heterogeneity, households and firms are instantiated as families of agents with rich variation in preferences, abilities, and other background, whereas the central bank and government are modeled as single institutional agents~\citep{Blanchard2025Convergence}. Agents are implemented through an LLM-based module that integrates environmental observations, traits, and structured memory to generate reasoning, planning, and decision-making. Following the agent–interaction–environment paradigm~\citep{Wooldridge2009MultiAgent}, we construct a simulation environment that features a frictional labor market, heterogeneous goods markets, and core financial interactions. The environment is augmented by a visualized map (Figure \ref{fig::pipeline}) to provide spatial context.
% that enhances the intuitive perception of geographic and economic dynamics.

Experimental results show that SimCity successfully exhibits a range of classic macroeconomic phenomena, such as Okun's law, the Phillips curve, the Beveridge curve, price elasticity of demand, and Engel's curve (detailed in Section~\ref{sec::experiments}). It also simulates the dynamic expansion of a city, and facilitates the simulation of novel economic shocks that were difficult to study in previous frameworks.

In summary, our work makes the following contributions:

1. \textbf{LLM-driven macroeconomic simulation with rich interactions}. We integrate LLMs into economic simulation by modeling four distinct economic roles, households, firms, a central bank, and a government, as LLM-driven agents. This enables agent heterogeneity and rich economic interactions.

2. \textbf{Visualized urban-style simulation environment}. We provide LLM-based agents with a mapped virtual city that provides spatial context and renders the dynamics of urban expansion, supporting decision-making in a realistic urban setting.

3. \textbf{Systematic evaluation}. We compile a checklist of canonical macroeconomic phenomena and regularities and use it to demonstrate that our framework exhibits established macroeconomic patterns. We further study responses to novel economic shocks, including exogenous price changes and shifts in production structure, highlighting our framework’s flexibility in analyzing scenarios that are difficult to capture in existing models.

% A. 统计搬家的频率和财产的关系
% B. 搬家的频率和经济形势（GDP增长率）的关系
% 找工作是否伴随搬家

% 假设技术变动对劳动上有影响（比如说信息化使得很多岗位需要新的工人）（假设足够自动化使得不需要某一类技能）

% 印一下 Gini 的曲线（比如说有可能经济更 productive 但反而会萎缩）

% some ideas might be looking at household mobility -- econagent for sure cannot do this. how frequent are people moving homes? are they related to when they change jobs?

% for another example, maybe we can look at shock to a specific section of the labor market, e.g. what if AGI is here and all service sector firms, broadly defined, no longer hire people -- what would happen to people with these skills, what would happen to the overall economy. 

% (if there is no tech/service sector in simcity, maybe shut down manufacturing sector? and have firms that rely on manufacturing sector goods get what they need at some low cost -- this is almost like modelling the shock to the US economy when China joined WTO)

% some other exercises can look at distributional consequences of taxation, which is something economists care a lot about. we already have Gini, so maybe look at GINI when taxation is super low or super high. 
\vspace{-2mm}
\section{Related Work}\label{sec::related_work}
\vspace{-2mm}
% \paragraph{LLMs for Social Simulation}
The application of LLMs to social simulation represents an emerging and promising research frontier. LLMs are capable of exhibiting human-like behaviors~\citep{xie2024largelanguagemodelagentsbehavior} and can be endowed with diverse personas~\citep{survey-chen2024persona}, offering a basis for interpretable agentic reasoning. Structurally, these simulation frameworks typically contain agents, an environment they interact with, and the interfaces that mediate their interactions~\citep{gao2024xiaochong}. Existing work in this domain can be broadly divided into two categories. The first category encompasses general-purpose social simulation frameworks such as ~\citet{park2023generativeagentsinteractivesimulacra},~\citet{tian2025visualizedframeworkeventcooperation},~\citet{tang2024gensim}, ~\citet{piao2025agentsocietylargescalesimulationllmdriven}, and ~\citet{huang2025adasocietyadaptiveenvironmentsocial}, which aim to model general societal patterns. The second category investigates specific social phenomena, such as public administration crises~\citep{xiao2023simulatingpublicadministrationcrisis}, health policy~\citep{hou2025societygenerativeagentssimulate}, political manipulations~\citep{touzel2024simulation}, financial market~\citep{gao2024simulatingfinancialmarketlarge}, and deduction games~\citep{xu2025languageagentsreinforcementlearning}.

\begin{wraptable}{r}{0.5\textwidth}
\vspace{-2mm}
 \centering
        \small
        \setlength{\tabcolsep}{2pt}
        \begin{tabular}{lccccc}
            \hline
            \textbf{Simulator} & \textbf{AI E.} & \textbf{LEN} & \textbf{CATS} & \textbf{EconA.} & \textbf{SimCity} \\ \hline 
            Households & 10 & 100 & 100 & 200 & 1000 \\
            % Tax Type & Labor & $\times$ & $\times$ & Labor & Labor \& VAT \\
            {Tax Type} & {Labor} & {\(\times\)} & {\(\times\)} & {Labor} & {Labor}\\
            {} & {} & {}& {} & {} & {\&VAT} \\ % Value-Added 太长了，建议缩写
            Agents' Types$^*$ & HG & HB & HFB & HBG & HFBG \\
            Markets$^\dagger$ & $\times$ & L & LGF & $\times$ & LGF \\
            Visualized Map & \checkmark & $\times$ & $\times$ & $\times$ & \checkmark \\
            Goods Type & 2 & 1 & 1 & 1 & 10+ \\ 
            Interpretability & $\times$ & $\times$ & $\times$ & \checkmark & \checkmark \\ \hline 
        \end{tabular}
        \vspace{2pt}
        \raggedright \scriptsize
        \textbf{Note:} VAT: Value-Added Tax. $^*$H: Household, F: Firm, B: Bank, G: Gov. $^\dagger$L: Labor, G: Goods, F: Financial.
        \caption{Comparison of features between frameworks listed in Section~\ref{sec::experiments}.}
        \label{table:comparison}
\end{wraptable}

% \paragraph{Traditional Macroeconomic Modeling}
DSGE and ABM  offer contrasting paradigms for macroeconomic modeling: DSGE emphasizes stylized optimizing agents and market-clearing equilibrium \citep{Stokey1989RecursiveMethods, Ljungqvist2018RecursiveMacro}, while ABM models heterogeneous agents with rule-based behaviors \citep{Tesfatsion2006HandbookACE}. While DSGE remains dominant \citep{Woodford2009Convergence, DelNegro2013FRBNYDSGE}, recent crises have renewed interest in ABMs \citep{Stiglitz2018ModernMacro, Borsos2024AgentBasedBoE}. Our framework combines key advantages of both approaches by enabling heterogeneous agents with flexible, LLM-based reasoning while preserving structured economic interactions.

%DSGE features highly idealized decision-makers characterized by precise mathematical objective functions, rational expectations about future variables, and strict market-clearing equilibrium conditions \citep{Stokey1989RecursiveMethods, Ljungqvist2018RecursiveMacro}. In contrast, ABM follows a bottom-up, simulation-driven methodology, where agents are endowed with rule-of-thumb behavior patterns and, in some cases, learning heuristics \citep{Tesfatsion2006HandbookACE, Axtell2025AgentBased}. Although DSGE remains the dominant approach in economics departments and policy institutions \citep{Woodford2009Convergence, DelNegro2013FRBNYDSGE, Chen2023EstimatedDSGE}, there has been a renewal of interest in ABMs, particularly following the Global Financial Crisis of 2008 and again after the COVID-19 crisis, as mainstream DSGE models struggled to account for these unprecedented shocks \citep{DelliGatti2008EmergentMacro, Stiglitz2018ModernMacro, Borsos2024AgentBasedBoE}. Our work is closer to the ABM tradition. However, we depart from conventional ABMs by replacing predetermined behavioral rules with LLM-powered agents capable of flexible reasoning and information processing in natural languages. 

% \paragraph{Closely Related Work} 
\cite{li2024econagentlargelanguagemodelempowered} introduces EconAgent as a recent attempt at LLM-driven macroeconomic simulation. Its framework restricts agent-environment interactions to two simplified variables: consumption share and work propensity, which constrains the range of economic activities that can be simulated. Our work extends this approach by incorporating other key economic elements, including firms as LLM-driven agents, heterogeneous goods markets, enriched financial activities, and a taxation system. Thus, our framework allows the exploration of a wider range of macroeconomic phenomena. Another contemporary work \citet{mi2025econgymscalableaitestbed} is proposed as a testbed for agent performance evaluation while we focus on urban development and macroeconomic laws.
\vspace{-2mm}
\section{Environment and Interaction}\label{sec::system}
\vspace{-2mm}
Our framework consists of three core layers: environment, interaction protocol, and agents. This section describes the environment and interaction of our system architecture.

\subsection{Environment}\label{sec::environment_layer}
On a high level, the simulation proceeds in two phases. In phase 1 (\textit{the move-in phase}), new households with synthetic profiles, representing immigrants or newborns, are introduced into the SimCity environment until a predetermined maximum population is reached. Unlike the typical burn-in period used purely for model stabilization in ABMs~\citep{lengnick2013agent}, this phase is explicitly designed to study urban expansion, capturing how economic activity and spatial development evolve as the city grows.

In phase 2 (\textit{the development phase}), the population is fixed while urban developments such as firm creation, investment, and spatial reallocation continue to occur. We design this phase to study macroeconomic regularities, as the system operates in 
% a quasi-steady state
a stable state
that is less susceptible to the transient noise present in the expansion period.

Phase 1 lasts 36 steps, and phase 2 lasts 144 steps in a typical simulation. Each simulation step represents one month. The order of agents' actions within each step is detailed in Appendix~\ref{app:stages}.
% Multi-thread is applied to generate response concurrently. Consequently,
Agents can only access information from previous steps.

There are three markets in the environment. Firms sell goods to each other and to households in a \textit{goods market}. Notably, goods are qualitatively differentiated (e.g., food, clothing). Firms post jobs and are matched with households in a \textit{labor market}. The central bank accepts deposits and provides loans in a \textit{financial market}. Details of environment setup can be found at Appendix~\ref{app:environment_setup}.

We develop a web-based render module to visualize the urban expansion as shown in Figure~\ref{fig::pipeline}. All buildings are displayed on the rendered map. The geographic placement of new firms is decided by a Visual-Language Model (VLM). For technical details of this module, please refer to Appendix~\ref{app:render_module}.

\subsection{Interaction Protocol}
\vspace{-2mm}
We leverage the common-sense reasoning capabilities of LLMs to act as human-like, heterogeneous agents. Details about the prompts used are provided in Appendix~\ref{app:prompting}. Agents interact with the environment by means of \textit{function calling}. The framework loads all the operations that the agent can execute and appends formatted function names along with their descriptions to the prompt. The LLM returns the actions to be taken and their parameters in JSON format. The framework will execute after a verification. For detailed examples, see Appendix~\ref{app:prompt_examples}.

%\vspace{-2mm}
%\paragraph{Function Calling}

% \footnote{Details about parameter calibration of the income distribution and the Input-Output Table can be found in the Appendix~\ref{app:agents_definition}}

\vspace{-2mm}
\section{Agents}\label{sec::modeling}
\vspace{-2mm}
% \begin{figure*}[t]
% \centering
%     \includegraphics[width=0.8\linewidth]{figs/paper.pdf}
% \caption{The illustration of our EconAgent~(left) and simulation environment~(right).}
% \label{fig:framework}
% \end{figure*}

This section presents an overview of the agents in SimCity. As illustrated in Figure~\ref{fig::pipeline}, there are four agent types: households, firms, a government, and a central bank, each simulated by an LLM. Building on prior work \citep{gatti2011macroeconomics,wolf2013multi,dawid2018agent,li2024econagentlargelanguagemodelempowered}, our enriched framework expands the decision space considerably. \begin{itemize}[noitemsep, topsep=0pt]
    \item Households make four key decisions each period: consumption bundle, labor market action, housing, and financial activity.
    \item Firms set production levels and prices, decide on hiring and capital investment, and acquire financing when needed.
    \item The government collects taxes and stimulates the economy through public spending and transfer payments to support consumption and improve social welfare.
    \item The central bank adjusts the interest rate in response to market conditions.
\end{itemize}
\vspace{-2mm}
\subsection{Households}
\textit{Households} are the fundamental units in our simulation and interact with all other agent types and markets. Each household is initialized with a heterogeneous profile, including age, education, consumption preference, and skill endowment. Details about these characteristics and the initialization process can be found in Appendix~\ref{app:profile_setup}. 

Each month, households receive two types of information: 
a personal report summarizing income, expenditures, and other status changes during the current period, and a citywide update containing goods prices, labor market conditions, housing availability, returns on investment, and interest rates. Additionally, households may receive offers for vacancies that align with their skills.

Based on this information, each household decides on (i) its consumption bundle, which requires a minimum expenditure on certain essential goods \citep{ravn2008macroeconomics}; (ii) its labor market action (accepting or rejecting a new job if offered, or resigning from its current job); 
(iii) its housing choice, and (iv) its financial decisions (saving, borrowing, or investing in the common investment pool, which is discussed below).

\subsection{Investment Pool and Firms}
\vspace{-2mm}
\paragraph{Investment Pool} An investment pool is implemented as an intermediary module that leverages a VLM for investment decisions. Funds from households’ investment actions are temporarily deposited into this pool and returned if unused for a month. When the pool accumulates sufficient capital and the VLM deems conditions favorable, a new firm is established. 

The pool chooses from a library of 44 synthetic firm templates, each of which produces a unique type of good, and selects a geographic placement for the firm. The construction of these templates from real-world data is detailed in Appendix~\ref{app:synthesis_firm}. After a firm is established, all contributing households receive shares of the firm in proportion to their investment.
\vspace{-2mm}
\paragraph{Firms} Each \textit{firm} in SimCity produces a single type of good. It hires households as workers, invests in capital to improve productivity, and transforms input goods into its specialized output, which is then sold in the market. Firms are instantiated from templates that specify skill requirements of job positions in the firm and the input-output relations governing its production process, where a firm may take possibly more than one input good for its production.
The effectiveness of resource utilization (i.e., the ratio at which input goods are transformed into the output good) is determined by a firm-specific Cobb-Douglas production function~\citep{cobbdouglars}:
\begin{equation}
    Y_i = A \, L_i^{1 - \alpha} \, K_i^\alpha,
\end{equation}
where \(A\) is total factor productivity, \(K_i\) is the i-th firm’s capital stock, \(L_i\) is the effective labor supply, defined as the amount of labor input adjusted for the match between job skill requirements and employees’ actual skills.

Similar to households, firms receive individual reports summarizing their internal operation (e.g., expenditure composition, employee skill profiles), and citywide updates on external conditions such as unemployment rate, interest rates, supply/demand conditions in the goods market. 

After deliberating on these information, each firm chooses (i) the quantity and price of its outputs, taking into consideration the prices of relevant input goods, as well as the geometric distance to their respective suppliers; (ii) its labor market actions, which may include posting job vacancies, laying off employees, or modifying wages; and (iii) its investment decision, including whether to borrow from the bank, and whether to purchase fixed assets to increase its capital.
%\(K\).

More detailed descriptions of firms can be found in Appendix~\ref{app:firm_detail}.
\vspace{-2mm}
\subsection{Government}
\vspace{-2mm}
\paragraph{Macroeconomic Indicators} The \textit{government} monitors key indicators about the state of the economy, and adjusts taxation and fiscal policies to enhance social welfare, as measured by these indicators. The most important macroeconomic indicators include total consumption, total investment, nominal GDP, and real GDP, where nominal GDP is calculated using current prices, and real GDP uses constant base-year prices. Inflation is tracked in two forms: wage inflation, defined as the rate of change in average wages, and GDP inflation, measured by the rate of change in the GDP deflator. The GDP deflator itself is the ratio of nominal GDP to real GDP. 
\vspace{-2mm}
\paragraph{Tax and Welfare}The government collects bracketed income tax from households and value-added tax (VAT) from firms. Let $\mathcal{B}=\{(b_k,r_k)\}_{k=1}^K$ denote a tax schedule, where $b_k$ is the lower threshold of bracket $k$, with $b_{K+1}=\infty$, and $r_k$ the corresponding marginal tax rate. The tax liability from agent \(i\) with tax base \(z_i\) (income for households, or profit for firms) is:
\begin{equation}
\label{eq:bracket-tax}
t_i
= \sum_{k=1}^{K}
  r_k\,
  \max\bigl(\,\min(\,z_i,\,b_{k+1}\,)-b_k, 0\, \bigr) .
\end{equation}
The total tax revenue collected by the government is the sum of taxes from all agents. The government uses this tax revenue in three ways: investing in the construction of public service buildings, which are modeled as government-owned firms, distributing it to households as a universal basic income (UBI), or reserving it for the future. 

% The fiscal policy, including the specifics of tax schedule \(\mathcal{B}\) and the usage of revenue, is adjusted at each step based on indicators perceived by the government agent.

Mathematical definitions of indicators and other details can be found in Appendix~\ref{app:government_detail}.
\vspace{-2mm}
\subsection{Central Bank and Financial System}
The \textit{central bank} accepts deposits, provides loans, and implements monetary policy. In the first month of each year, it sets the policy interest rate according to a modified Taylor rule, which is widely used in monetary policies.~\citep{GaliGertler2007Macroeconomic,dawid2018agent}. Formally:
\begin{equation}
\label{eq:taylor_rule}
\hat{r} = \max\bigl(r_\text{natural} + \pi_\text{target} + \alpha (\pi - \pi_\text{target}) + \beta (Y - Y_\text{natural}),\, 0\bigr),
\end{equation}
where \(\hat{r}\) is the policy rate set by the central bank, \(r_\text{natural}\) is the long-run natural interest rate, \(\pi_\text{target}\) is the target inflation rate, \(\pi\) is the GDP inflation in the last period, \(Y\) is the actual output (GDP), and \(Y_\text{natural}\) is the potential output as measured by a linear trend. Parameters \(\alpha\) and \(\beta\) capture the central bank's responsiveness to inflation and output gap, respectively. 

Intuitively, when inflation is high, or output exceeds its long-term trend (signals of an overheated economy), the central bank raises interest rates to cool the economy and maintain monetary stability, and vice versa \citep{Gali2015NewKeynesian}. To reflect the gradual adjustment observed in real-world monetary policy, we incorporate a smoothing term consistent with empirical evidence on persistent rate changes \citep{10.1257/mac.4.4.126}.

At each time step, deposits and loans accrue interest: the deposit rate equals the policy interest rate, while the loan rate equals the policy rate plus a fixed markup. Further details about the financial system are in Appendix \ref{app:cb}.

% When considering the dynamics of labor, consumption, and financial markets, the influence of these macroeconomic trends on agent decision-making is also seldom considered, raising the second challenge. 

\vspace{10mm}
\section{Experiments}\label{sec::experiments}
\vspace{-2mm}

\begin{wraptable}{r}{0.5\textwidth}
\centering
\small
\setlength{\tabcolsep}{2pt}
\begin{tabular}{lccccc}
    \hline
    \textbf{Regularity} & \textbf{AI-E.} & \textbf{LEN} & \textbf{CATS} & \textbf{EconA.} & \textbf{SimCity} \\ \hline
    Phillips Curve   & / & $\times$ & $\times$ & \checkmark & \checkmark \\
    Okun's Law       & / & \checkmark & \checkmark & \checkmark & \checkmark \\
    Beveridge Curve  & / & $\times$\textsuperscript{\dag} & \checkmark & / & \checkmark \\
    Price Elasticity & / & / & / & / & \checkmark \\
    Engel's Law      & / & / & / & / & \checkmark \\
    Invest. Volat.   & / & / & / & / & \checkmark \\
    % Price Stickiness & / & / & / & / & \checkmark \\ 
    \hline
\end{tabular}
\vspace{2pt}
\raggedright \scriptsize
\textbf{Legend:} \checkmark: Verified; $\times$: Inconsistent; /: Not available. 
\caption{Macroeconomic regularities exhibited by SimCity, details are in Section~\ref{sec::experiments}.}
\label{table:macro_phenomena_comparison}
\end{wraptable}

{
  \renewcommand{\thefootnote}{}
  \footnotetext{\textsuperscript{\dag}The paper by \cite{lengnick2013agent} claims to have verified the Beveridge Curve, but we are unable to reproduce it with \url{https://github.com/newwayland/baseline-economy}}
}

\begin{figure}[t]
\vspace{-2mm}
\centering
    \begin{subfigure}[b]{0.49\linewidth}
        \includegraphics[width=\linewidth]{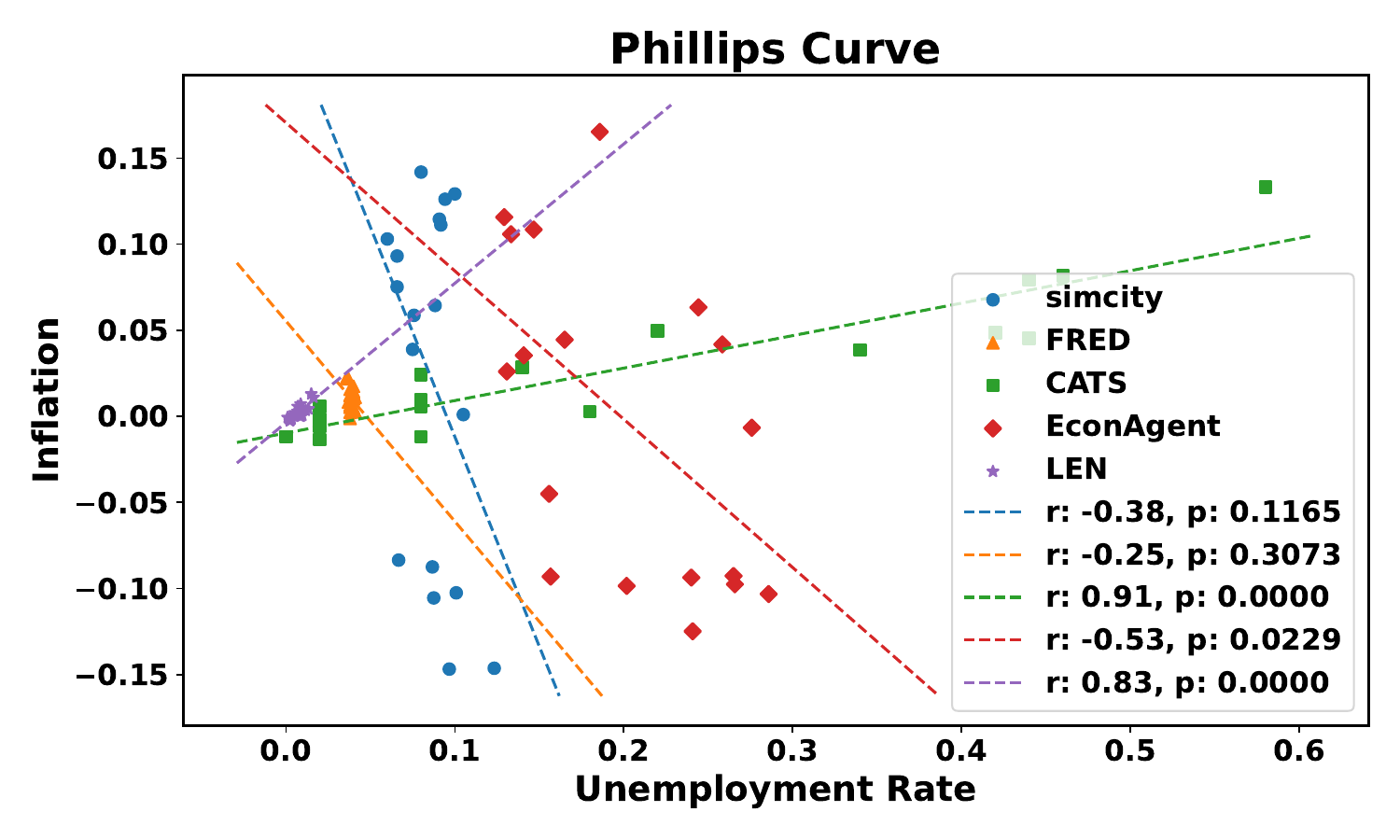}
    \end{subfigure}
    \vspace{-1mm}
    \begin{subfigure}[b]{0.49\linewidth}
        \includegraphics[width=\linewidth]{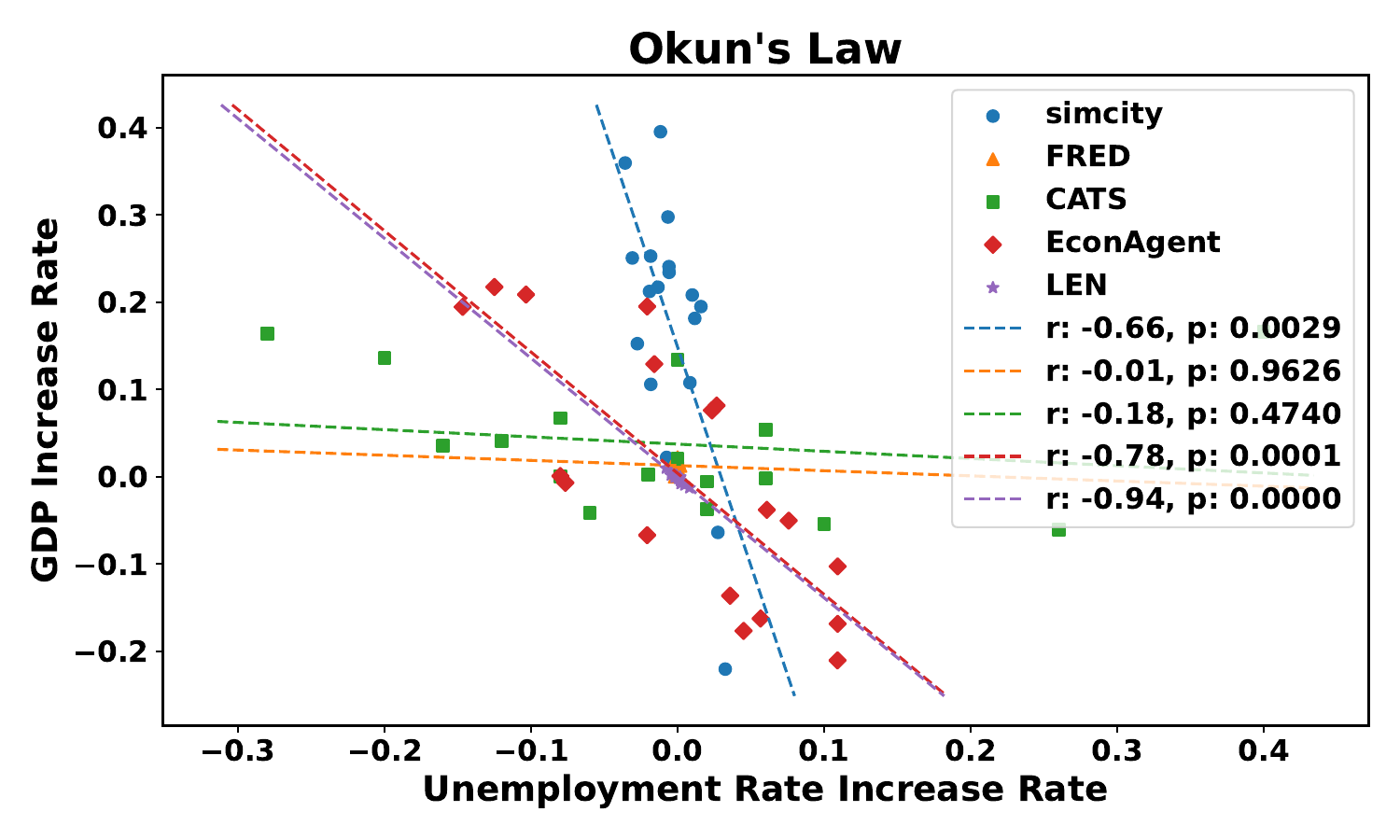}
    \end{subfigure}
    \caption{Emergence of the Phillips Curve and Okun’s Law in SimCity simulations. $r$-value is the Pearson correlation coefficient, and $p$-value indicates the statistical significance of it.}
    \label{fig::classic-curves}
\end{figure}

We conduct experiments to evaluate SimCity's ability to perform macroeconomic simulations and to display urban development dynamics. A key limitation of existing frameworks~\citep{lengnick2013agent, li2024econagentlargelanguagemodelempowered} is the lack of a common basis for comparison. Prior studies often highlight a relatively small set of selected macroeconomic phenomena, making it difficult to gauge and compare the capacities of different simulation frameworks.

To address this gap, we compile a checklist of canonical macroeconomic ``stylized facts'' from the broad economics literature \citep{blanchard1989lectures, williamson2014macroeconomics, jorda2017macrofinancial, Axtell2025AgentBased}, many of which have previously been used to evaluate other ABM models.
We then use this checklist to assess SimCity's ability to exhibit established patterns and to simulate new ones under novel economic shocks. Taking advantage of SimCity's rich framework, we aim to answer the following research questions:
% \begin{itemize}
%     \item \textbf{RQ1}: What phenomena emerge in SimCity, compared with prior simulation environments?
%     \item \textbf{RQ2}: Are emergent regularities robust across multiple independent simulations?
%     \item \textbf{RQ3}: Can SimCity exhibit realistic spatial patterns during the move-in and urban expansion phase?
%     \item \textbf{RQ4}: To what extent can SimCity respond to external shocks?
%     % Can the simulation based on SimCity reflect the impact of external intervention?
%     % \item \paragraph{RQ4}: What do different LLMs 
%     %, as visualized by the VLM
% \end{itemize}

\textbf{RQ1}: What phenomena emerge in SimCity, compared with prior simulation environments?

\textbf{RQ2}: Are emergent regularities robust across multiple independent simulations?

\textbf{RQ3}: Can SimCity exhibit realistic spatial patterns during the move-in and urban expansion phase?

\textbf{RQ4}: To what extent can SimCity respond to external shocks?
    % Can the simulation based on SimCity reflect the impact of external intervention?
    % \item \paragraph{RQ4}: What do different LLMs 
    %, as visualized by the VLM
\vspace{-2mm}
\subsection{Experiment Setup}
\vspace{-2mm}
We simulate an economy with a maximum of 1000 households. Each simulation proceeds in two phases. During the initial phase, which lasts 36 steps (3 years), new households gradually populate the city. During the second phase, which lasts 144 steps (12 years), the population remains fixed, and we observe the simulation under steady-state conditions. We use \texttt{gpt-4o-mini} for regular reasoning and \texttt{gpt-4} as the vision language model for establishment decision-making. Both are provided by Azure OpenAI API\footnote{\url{https://portal.azure.com/}} with default sampling parameters. We provide detailed examples of prompts and responses in Appendix~\ref{app:prompt_examples}. We also explore the robustness of SimCity across alternative LLMs in Appendix~\ref{app:robustness}.

For comparison, we select two traditional ABMs with predetermined rules LEN~\citep{lengnick2013agent} and CATS~\citep{gatti2011macroeconomics}, a deep multiagent reinforcement learning model AI-economist~\citep{zheng2022ai}, a LLM-based system EconAgent~\citep{li2024econagentlargelanguagemodelempowered}, and real-world data FRED from 1970 Q1 (Federal Reserve Economic Data)\footnote{\url{https://fred.stlouisfed.org/}}.

\subsection{Macroeconomic Emergence (RQ1)}\label{sec::rq1}

A central test of macroeconomic simulations is its ability to exhibit well-documented empirical regularities (``stylized facts'') observed in real-world economies.
Leveraging the strong role-play capabilities of LLM agents, SimCity captures many of these regularities that traditional agent-based models have historically struggled to generate.
Table~\ref{table:macro_phenomena_comparison} summarizes the key phenomena and compares SimCity’s performance with baseline models.

\begin{figure*}[t]
\vspace{-2mm}
    \centering
    \includegraphics[width=\linewidth]{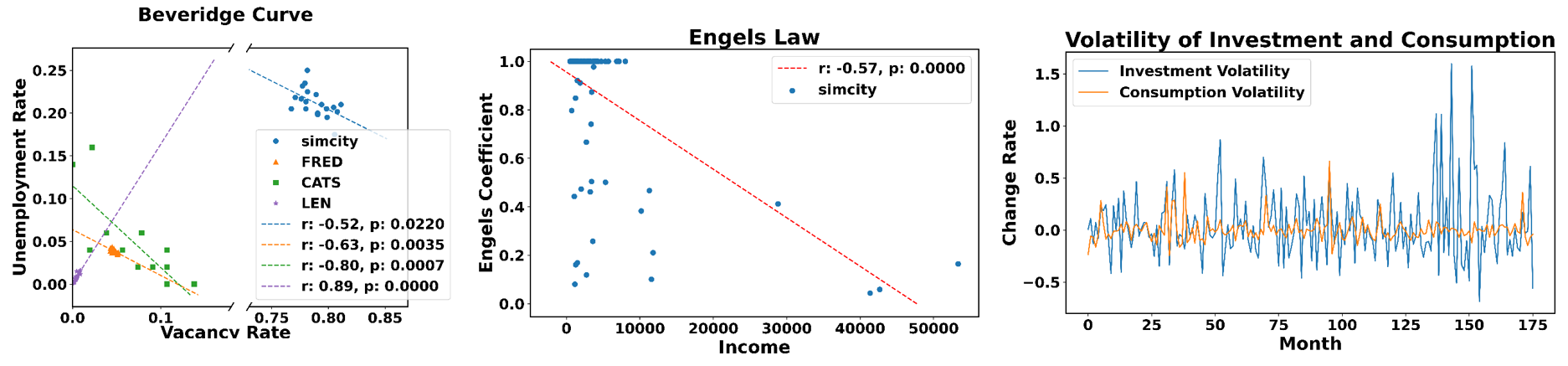}
    \caption{Macroeconomic emergences from SimCity: Beveridge Curve, Engels Law, and business cycle volatility. Business cycle volatility refers to the well-established phenomenon that investment volatility is significantly higher than consumer spending volatility across economic fluctuation~\citep{10.1257/aer.97.3.586}.}
\label{fig::new-curves}
\label{fig::emergences}
\end{figure*}

% \begin{figure}[t]
%     \centering
%     \def\graphheight{3cm} 
%     \begin{subfigure}[b]{0.3\textwidth}
%         \centering
%         \includegraphics[height=\graphheight]{Fig/beveridge_curve_broken_axis_comparison.pdf}
%         \label{fig::price_down_impulse}
%     \end{subfigure}
%     \begin{subfigure}[b]{0.35\textwidth}
%         \centering
%         \includegraphics[height=\graphheight]{Fig/engels_law_plot.pdf}
%         \label{fig::skill_impulse}
%     \end{subfigure}
%     \begin{subfigure}[b]{0.3\textwidth}
%         \centering
%         \includegraphics[height=\graphheight]{Fig/investment_volatility.pdf}
%         \label{fig::skill_impulse}
%     \end{subfigure}
%     \caption{Experimental results of exogenous impulses in SimCity. (a) A skill-replacement shock leads to an income gap between IT-proficient agents and others. (b) An external price impulse at year 15 shows that prices maintain their long-term tendency, demonstrating price stickiness.}
% \label{fig::emergences}
% \end{figure}

\vspace{-2mm}
\paragraph{Phillips Curve and Okun's Law}
The Phillips Curve~\citep{phelps1967phillips} describes the short-run inverse relationship between unemployment and inflation, while Okun's Law~\citep{okun1963potential} depicts the quarterly negative relationship between changes in the unemployment rate and real GDP growth. Following prior work  \citep{li2024econagentlargelanguagemodelempowered}, we use these well-known empirical regularities as major tests of the plausibility of macroeconomic dynamics in our simulation. Recent economic studies using more modern methodologies and granular data document that the Phillips Curve slope is small and has flattened in recent decades \citep{Hazell2022PhillipsSlope, Furlanetto2024PhillipsSlope}. Consequently, a steeper or more negative slope does not necessarily indicate a more faithful reproduction of the Phillips Curve. We plot the relationships using one data point per quarter in Figure~\ref{fig::classic-curves}\footnote{For simulations that operate on an annual time scale, we use annual data points instead.}. As shown in Figure~\ref{fig::classic-curves}, SimCity successfully demonstrates these relationships.\footnote{We omit AI-Economist from this comparison because unemployment is not incorporated in its framework.}  
% We acknowledge that the SimCity points have a large $p$-value (this could be improved by a larger sample size), but we empirically verify that the negative slope can be reproduced with different random seeds in Section~\ref{sec::robustness}.

% \begin{figure}[hbtp]
% \centering
% \begin{minipage}{\textwidth}
% \begin{subfigure}[b]{0.33\textwidth}
% \includegraphics[width=\textwidth]{Fig/beveridge_curve_broken_axis_comparison.pdf}
% \end{subfigure}
% \begin{subfigure}[b]{0.33\textwidth}
% \includegraphics[width=\textwidth]{Fig/engels_law_plot.pdf}
% \end{subfigure}
% \begin{subfigure}[b]{0.33\textwidth}
% \includegraphics[width=\textwidth]{Fig/investment_volatility.pdf}
% \end{subfigure}
% \end{minipage}
% \begin{minipage}{\textwidth}
% \centering
% \begin{subfigure}[b]{0.40\textwidth}
% \includegraphics[width=\textwidth]{Fig/elasticity.pdf}
% \end{subfigure}
% \begin{subfigure}[t]{0.4\textwidth}
% \includegraphics[width=\textwidth]{Fig/elasticity_legend.pdf}
% \end{subfigure}
% \end{minipage}
% \caption{SimCity also demonstrates phenomena such as the Beveridge Curve, which previous works cannot present.} \label{fig::new-curves}
% \label{fig::elasticity}
% \end{figure}

\vspace{-2mm}
\paragraph{Beveridge Curve} The Beveridge Curve captures the negative relationship between the job vacancy rate, which is the number of unfilled job openings as a proportion of total job postings, and the unemployment rate~\citep{Blanchard1989BeveridgeCurve}.  SimCity successfully exhibits this relationship, whereas the previous LLM-driven ABM, EconAgent, lacks the firm module needed to model job vacancies.

\vspace{-2mm}
\paragraph{Engel’s Law} Engel's Law states that the proportion of income spent on food (the Engel coefficient) decreases as income rises \citep{chai2010engel}. Simulations from SimCity correctly exhibit this pattern, reflecting that LLM-driven agents display human-like preferences.
Notably, validating Engel’s Law requires a heterogeneous goods market, which previous frameworks lacked and were therefore unable to study.

% \begin{figure}[t]
% \centering
% \begin{minipage}{\textwidth}
% \begin{subfigure}[b]{0.25\textwidth}
% \includegraphics[width=\textwidth]{Fig/elasticity.pdf}
% \end{subfigure}
% \begin{subfigure}[b]{0.25\textwidth}
% \includegraphics[width=\textwidth]{Fig/elasticity_legend.pdf}
% \end{subfigure}
% \end{minipage}
% \caption{Estimated Price Elasticity of Demand (PED) across different product categories in the SimCity environment, obeying the Law of Demand and commonsense notions of necessity.} \label{fig::elasticity}
% \end{figure}

\begin{wrapfigure}{r}{0.4\textwidth}
\begin{center}
\includegraphics[width=0.38\textwidth]{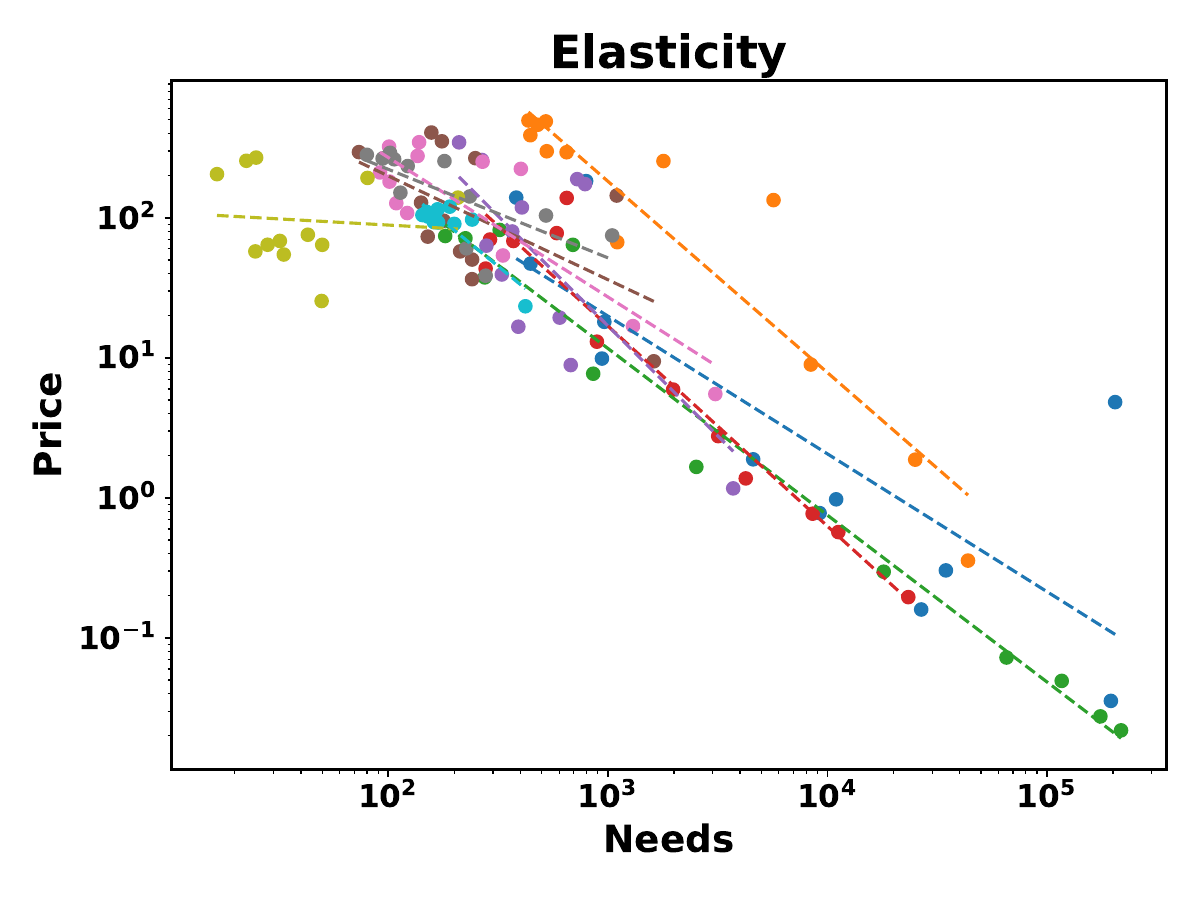}
\includegraphics[width=0.38\textwidth]{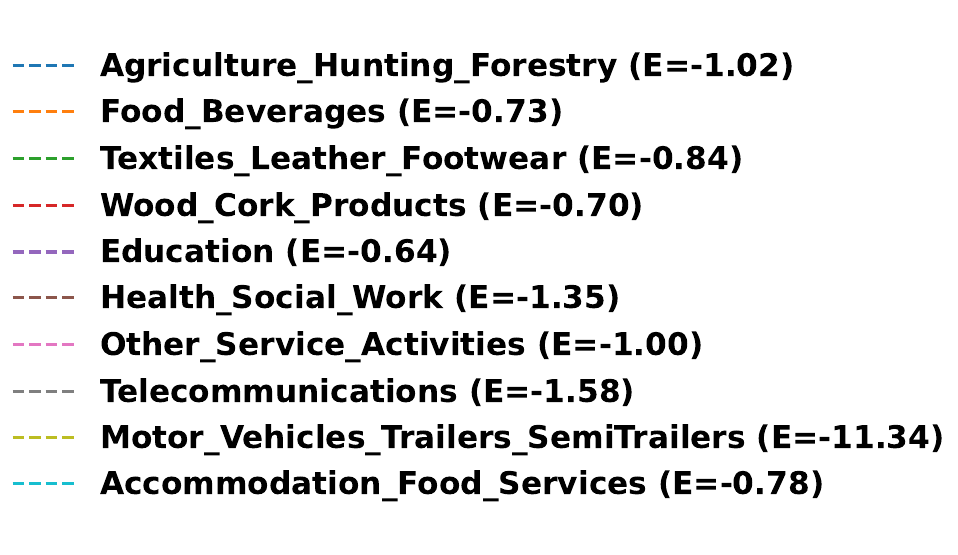}
\caption{Estimated Price Elasticity of Demand (PED) across different product categories in the SimCity environment, obeying the Law of Demand and commonsense notions of necessity. \( E = \frac{\Delta\log Q}{\Delta \log P} \) is the elasticity defined in Section~\ref{sec::rq1}.} \label{fig::elasticity}
\end{center}\vspace{-2mm}
\end{wrapfigure}

% \vspace{-2mm}
\paragraph{Law of Demand} To test the Law of Demand, we estimate the Price Elasticity of Demand (PED) for each good, defined as \( E = \frac{\Delta\log Q}{\Delta \log P} \), which measures how quantity demand responds to changes in price. Standard economic theory predicts a negative elasticity for normal goods, but the magnitude depends on the importance of the good to consumers~\citep{MasColell1995MicroeconomicTheory}. As shown in Figure~\ref{fig::emergences}, SimCity not only exhibits the general negative association predicted by the Law of Demand but also captures the variation in elasticity across goods. For example, necessities like ``food/beverages'' are more inelastic (\(-1<E<0\)), whereas goods such as ``motor/vehicles/trailers/semitrailers'' are more elastic (\(E<-1\)), which is a pattern confirmed by numerous empirical studies~\citep{perloff2009microeconomics_elasticity,elasticity_nealson2014_motor}. Lastly, as shown in Figure~\ref{fig::new-curves} (top-right), our volatility of investment is larger than that of consumption, which is consistent with real-world business cycle patterns.
% \vspace{-1mm}
\subsection{Simulation Robustness (RQ2)}\label{sec::robustness}
 % \vspace{-1mm}
An important question for LLM-driven simulations is whether the observed economic phenomena remain stable and reproducible across runs. To address this, we run three sets of experiments with the same hyperparameters but different random seeds, and the results demonstrate that the observed regularities are robust. Experiment results can be found at Figure~\ref{fig::stability-curves} in Appendix~\ref{app:robustness}.

\begin{figure*}[t]
\vspace{-2mm}
\centering
\includegraphics[width=\textwidth]{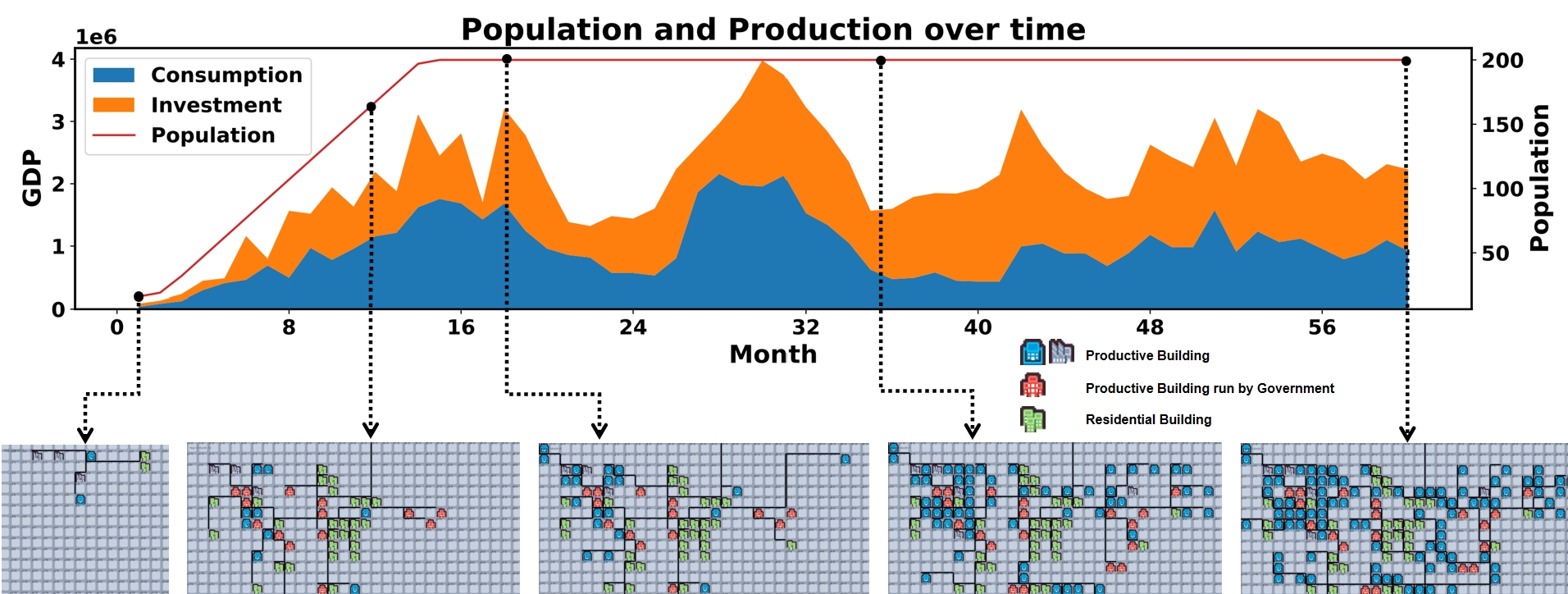}
\caption{GDP \& population curves, and map changes during the move-in phase. Another example is given in Appendix~\ref{app:extra_example_city_expansion}.}
\label{fig::phase1}
\end{figure*}

\subsection{Urban Expansion During Move-In (RQ3)}\label{sec::phase1}
As mentioned in Section~\ref{sec::environment_layer}, our simulation involves a phase in which households move in. Figure~\ref{fig::phase1} illustrates the developments during this stage. As shown in the figure, in the first 15 months, GDP steadily increases with the influx of residents. Then, as the influx ceases, the city transitions from a period of expansion to one of stable development after more than ten months of fluctuations. Meanwhile, we note that the VLM, without additional prompts, forms a clustered structure with residential areas near the center and production buildings on the periphery, in line with Alonso-Muth-Mills (AMM) model of the mono-centric city \citep{brueckner2011lectures}.

To evaluate the role of visual information in generating realistic urban forms, we conduct an ablation study to evaluate the necessity of the visual module by replacing the map images with text-based descriptions of the city layout. Experiments show that without visual grounding, the city evolves into unrealistic geometries, such as narrow linear formations, contrasting with the organic clusters formed by the VLM-empowered agents (visual comparison provided in Appendix~\ref{app:visual_module_ablation}).

% \begin{figure}[t]
%     \centering
%     \def\graphheight{3cm} 
%     \begin{subfigure}[b]{0.68\textwidth}
%         \centering
%         \includegraphics[height=\graphheight]{Fig/price_down_impulse.pdf}
%         \includegraphics[height=3cm]{Fig/price_down_impulse_legend.pdf} 
%         \caption{Impact on prices.}
%         \label{fig::price_down_impulse}
%     \end{subfigure}
%     \begin{subfigure}[b]{0.3\textwidth}
%         \centering
%         \includegraphics[height=\graphheight]{Fig/skill_impulse_gap_response_2.pdf}
%         \caption{Impact on skill.}
%         \label{fig::skill_impulse}
%     \end{subfigure}

%     \caption{Experimental results of exogenous impulses in SimCity. (a) An external price impulse at year 15 shows that prices maintain their long-term tendency, demonstrating price stickiness. (b) A skill-replacement shock leads to an income gap between IT-proficient agents and others.}
%     \label{fig::impulses_combined}
% \end{figure}

\subsection{Responses to Exogenous Shocks (RQ4)}\label{sec::price_impulse}
\paragraph{Price shock} To investigate whether our framework can be used to analyze the effects of sudden external shocks to the economy, we separately apply a price-down and a price-up impact on the city. At the beginning of year 15, we randomly select 7 of 44 goods in two different simulations and apply a 50\% price-down and a price-up impulse separately. We then let the simulation run for 6 years. Conceptually, one can think of these sudden, ephemeral changes in price levels of the selected goods as coming from exogenous shocks in the world market. Take electricity prices as an example, which is one of the shock goods in our simulations; it could come from a significant but brief conflict among energy-producing countries. 

\begin{wrapfigure}{r}{0.42\textwidth}
\begin{center}
\includegraphics[width=0.4\textwidth]{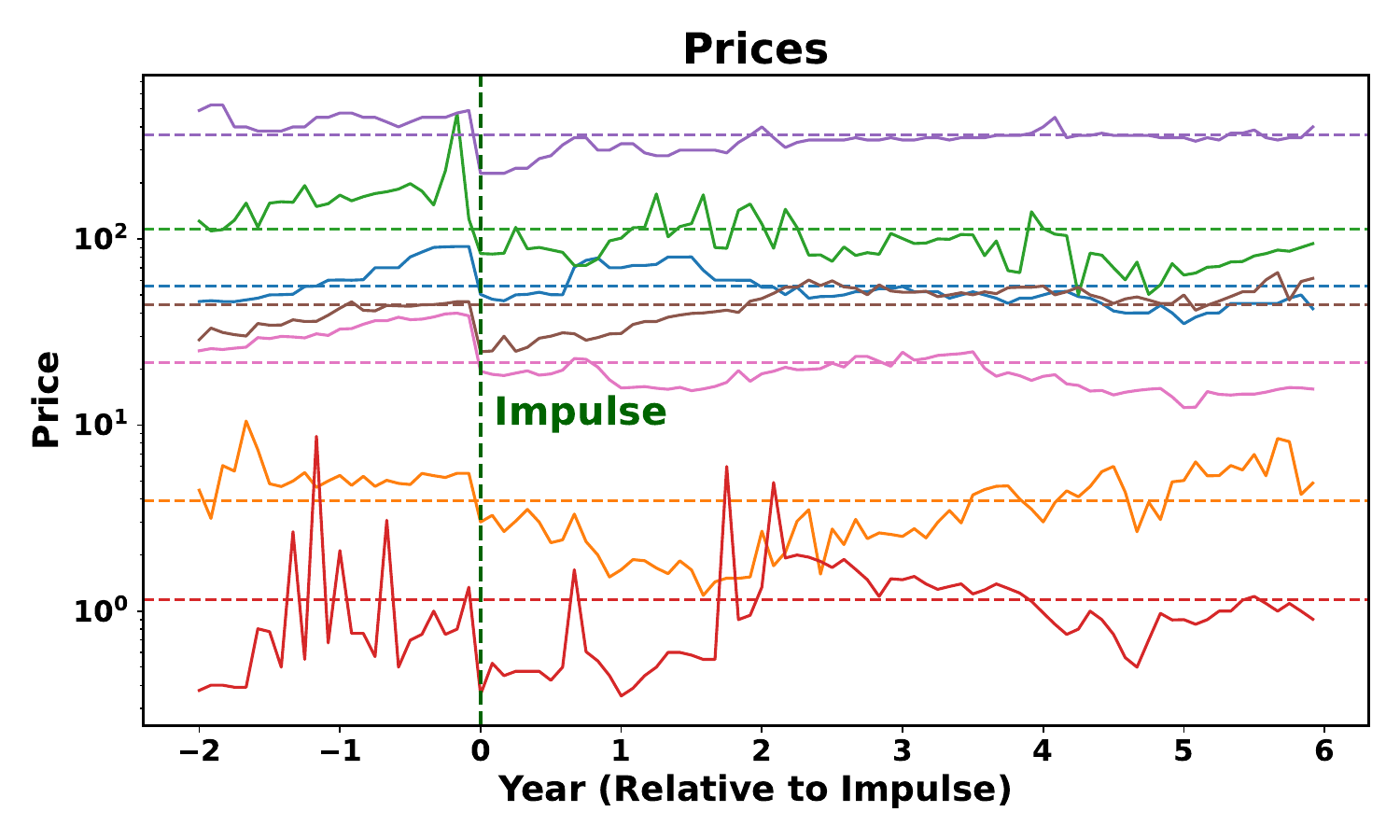}
\vspace{-2mm}
\includegraphics[width=0.4\textwidth]{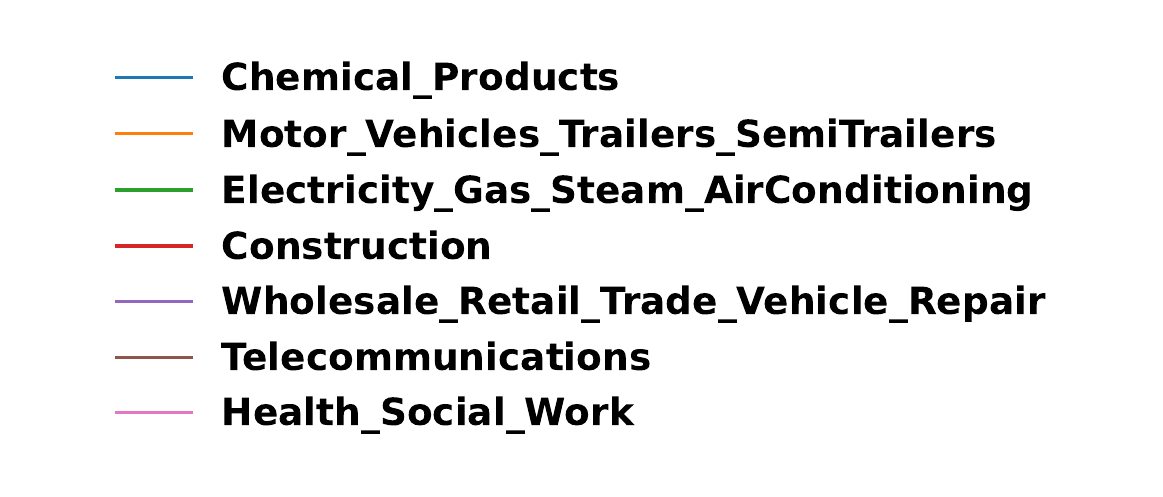}
\caption{An external price impulse at year 15 shows that prices maintain their long-term tendency, demonstrating price stickiness.}
\label{fig::price_down_impulse}
\end{center}
\end{wrapfigure}

Impulse responses to price-down shock are shown as Figure \ref{fig::price_down_impulse} and price-up at Figure \ref{fig::price_up_impulse}~(Appendix~\ref{app:price_impulse}). The experiment shows that the good's price level gradually returns to equilibrium after an exogenous shock. Due to price stickiness, the full effects of the exogenous shock take some time to develop. Importantly, note that dotted lines in the figure represent long-run average prices of goods. Since the shock is one-off and the economy's fundamentals did not change, prices revert to their original levels in the long run, despite the temporary dislocation. This stylized experiment showcases SimCity's ability to produce qualitative predictions in novel economic scenarios. 

\vspace{-2mm}
%is determined by its endogenous equilibrium and
% impact and display their response again

\begin{wrapfigure}{r}{0.42\textwidth}
\begin{center}
\includegraphics[width=0.4\textwidth]{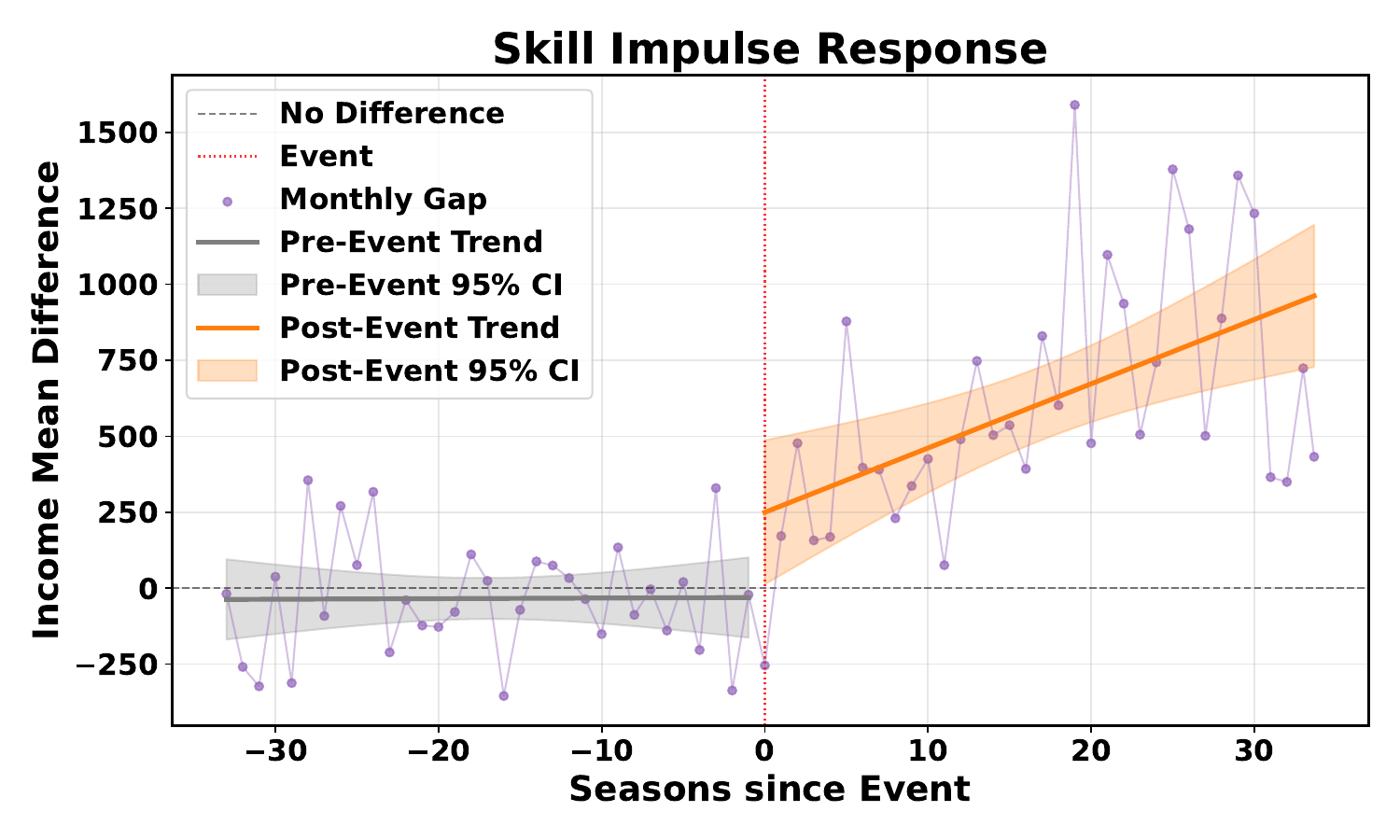}
% \caption{A skill-replacement shock leads to an income gap between IT-proficient agents and others.}
\caption{An automaton shock leads to an income gap between benefited agents and others.}
\label{fig::skill_impulse}
\vspace{-2mm}
\end{center}
\end{wrapfigure}

\paragraph{Automation Shock} To further demonstrate the ability of our framework to simulate novel economic scenarios, we consider a stylized automation shock where physical labor is entirely replaced by information technology (IT) skills (e.g., drivers become obsolete as cars become fully autonomous). We implement this shock by altering the skill requirements in production midway through the simulation (denoted as “0 to Event” in Figure~\ref{fig::skill_impulse}). As Figure~\ref{fig::skill_impulse} shows, following the shock, a substantial and widening income gap emerges between workers with strong labor skills and those with strong IT skills. This divergence persists as the economy expands, highlighting the distributional consequences of skill-biased technological change. Shaded areas represent 95\% confidence intervals.

% \begin{figure}[hbtp]
% \centering
% \begin{subfigure}[b]{0.49\textwidth}
% \includegraphics[width=\textwidth]{Fig/skill_impulse_gap_response_2.pdf}
% \end{subfigure}
% \caption{Impact of exogenous skill-replacement shock on income distribution. The purple line represents the evolving income gap between IT-proficient agents and other workers.} \label{fig::skill_impulse}
% \end{figure}

% \begin{figure}[h]
% \centering
% \begin{minipage}{\textwidth}
% \begin{subfigure}[b]{0.30\textwidth}
% \includegraphics[width=\textwidth]{Fig/price_down_impulse.pdf}
% \end{subfigure}
% \begin{subfigure}[b]{0.20\textwidth}
% \includegraphics[width=\textwidth]{Fig/price_down_impulse_legend.pdf}
% \end{subfigure}
% \end{minipage}
% \caption{External impulse (at year 15) does not significantly affect the tendency of prices of goods.} \label{fig::price_down_impulse}
% \end{figure}

\section{Limitations}
% \paragraph{Humanlike Agent} 我们使用大语言模型为 agent 制订决策，因此无法为大语言模型的决策施加足够强的约束，因为大语言模型不会真的像人类一样需要吃饭睡觉；决策可能缺乏 consistency；大语言模型会以一个非常微小的概率作出一些非常偏离预期的决策（例如在决定价格的时候直接调整数倍），而在近万次调用中这样的决策的概率不能忽略不计，这迫使我们给大语言模型的行为加上一些经验性的检验。
% \paragraph{更加细致的模拟} 我们的模拟框架中没有模拟现实中的债券、股票等市场以及衍生品市场，投资者也没有使用数学方法最优化他们的行为；这可能使得市场无法良好地模拟

\vspace{-2mm}
\paragraph{Abnormal Behavior from Agents.} Our agent's decision-making is powered by a LLM. A significant challenge arises from the inherent difficulty in imposing robust constraints on the LLMs' decisions. Furthermore, there exists a non-trivial probability that the LLM may generate highly aberrant or unexpected decisions, such as altering a price by several orders of magnitude. While the likelihood of any single such event is small, its probability becomes non-negligible over hundreds of iterations. In this work, we apply simple heuristic-based checks to prevent such egregious behavior.

\vspace{-2mm}
%Unlike humans, the LLM does not operate under real-world necessities such as eating or sleeping, which can lead to a lack of consistency in its choices. 
\paragraph{Lack of Complex Financial Activities} Our current simulation framework does not incorporate real-world financial markets, such as those for bonds, stocks, or their derivatives. Moreover, the simulated agents do not employ mathematical optimization techniques in their investing strategies. These simplifications may limit the model's ability to simulate complex market dynamics accurately.
\vspace{-2mm}

\section{Conclusion}\label{sec::conclusion}
%In this work, we explore the potential of LLMs to play as an agent in Macroeconomic Simulation and construct a comprehensive simulation framework. We design a checklist of key macroeconomic phenomena, and experiments show that . Our work prove that our framework can serve as a powerful new tool for economic analysis and policy evaluation. 

With SimCity, we conduct complex simulations that incorporate four distinct LLM-driven economic roles (households, firms, the central bank, and the government), with a visualized map and markets with heterogeneous goods. Experiments show that SimCity robustly matches a checklist of established macroeconomic phenomena. Moreover, when applied to novel economic scenarios, SimCity generates economically coherent and informative predictions. We hope SimCity could serve as a solid base platform for a more realistic economic simulation.

% However, in some phenomena, the simulation results of this framework still differ from theoretical expectations, and some modules are still overly simplified. These limitations await improvement, breakthroughs, and exploration by future researchers.

\vspace{-2mm}
\section{Ethics Statement}

This work utilizes LLM-driven agents to simulate. No sensitive or private data is used. Names of agents are generated by LLMs and are not intended to represent any real individuals or entities. Our simulation has not been quantitatively calibrated with real-world data and does not constitute a prediction of reality.

% \ificlrfinal
\section*{Acknowledgement}

We sincerely thank Jian Li, Kaifeng Lyu, and Boyuan Zheng for their valuable discussions and insightful comments that help improve this work.
% \fi
% \newpage
% , potentially limiting the applicability and generalizability of our results to other countries."
\vspace{-2mm}
% \section{Reproducibility Statement}
% To help with the reproducibility of our work, we provide all LLMs prompts in Appendix~\ref{app:prompting}. We call LLMs API with default sampling parameters. We repeat experiments with different random seeds (Appendix~\ref{app:robustness}) to ensure that our main results are robust.
% We plan to release code in the near future. 

\bibliography{colm2026_conference}
\bibliographystyle{colm2026_conference}

\appendix
\newpage

\section{Statement on LLM Usage}
This work utilizes LLM-driven agents to simulate. Additionally, we use LLMs to improve the grammar and readability of this manuscript. The LLM was not used for any other scientific aspects of this work, and all intellectual content is solely the product of the authors.

\section{Experiment Detail And Additional Results}

\paragraph{LLMs Costs} Our simulation costs about 800,000 tokens, which is roughly \$0.25 per step. For a standard 180-step simulation, the total cost is around \$180.

\paragraph{Base Year} We set prices for the first year as the base-year prices for calculating real GDP.

\subsection{Robustness}\label{app:robustness}

We run three sets of experiments with the same hyperparameters but different random seeds, and the results demonstrate that the observed regularity is reproducible.

\begin{figure*}[hbtp]
\centering
\begin{minipage}{\textwidth}
\begin{subfigure}[b]{0.31\textwidth}
\includegraphics[width=\textwidth]{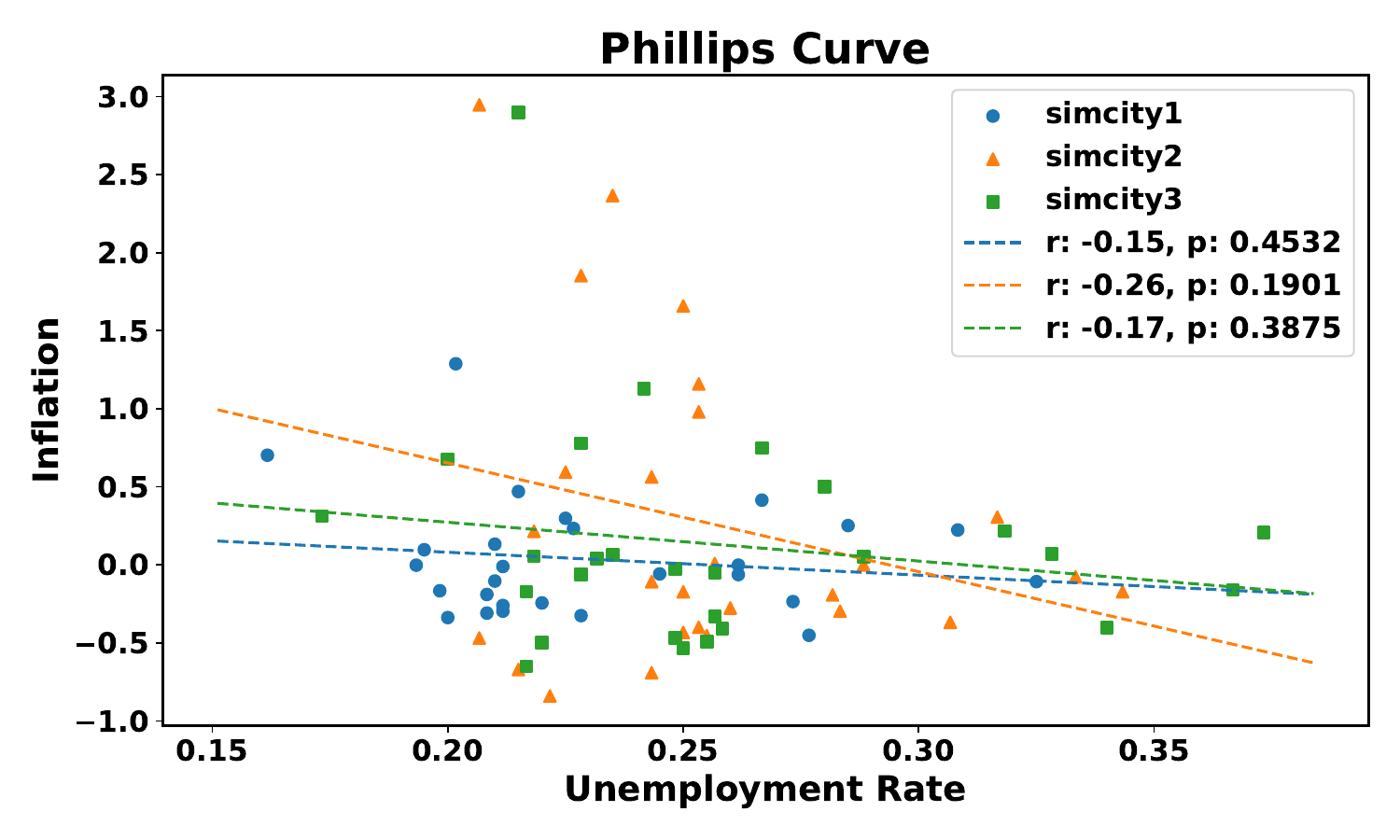}
\end{subfigure}
\begin{subfigure}[b]{0.31\textwidth}
\includegraphics[width=\textwidth]{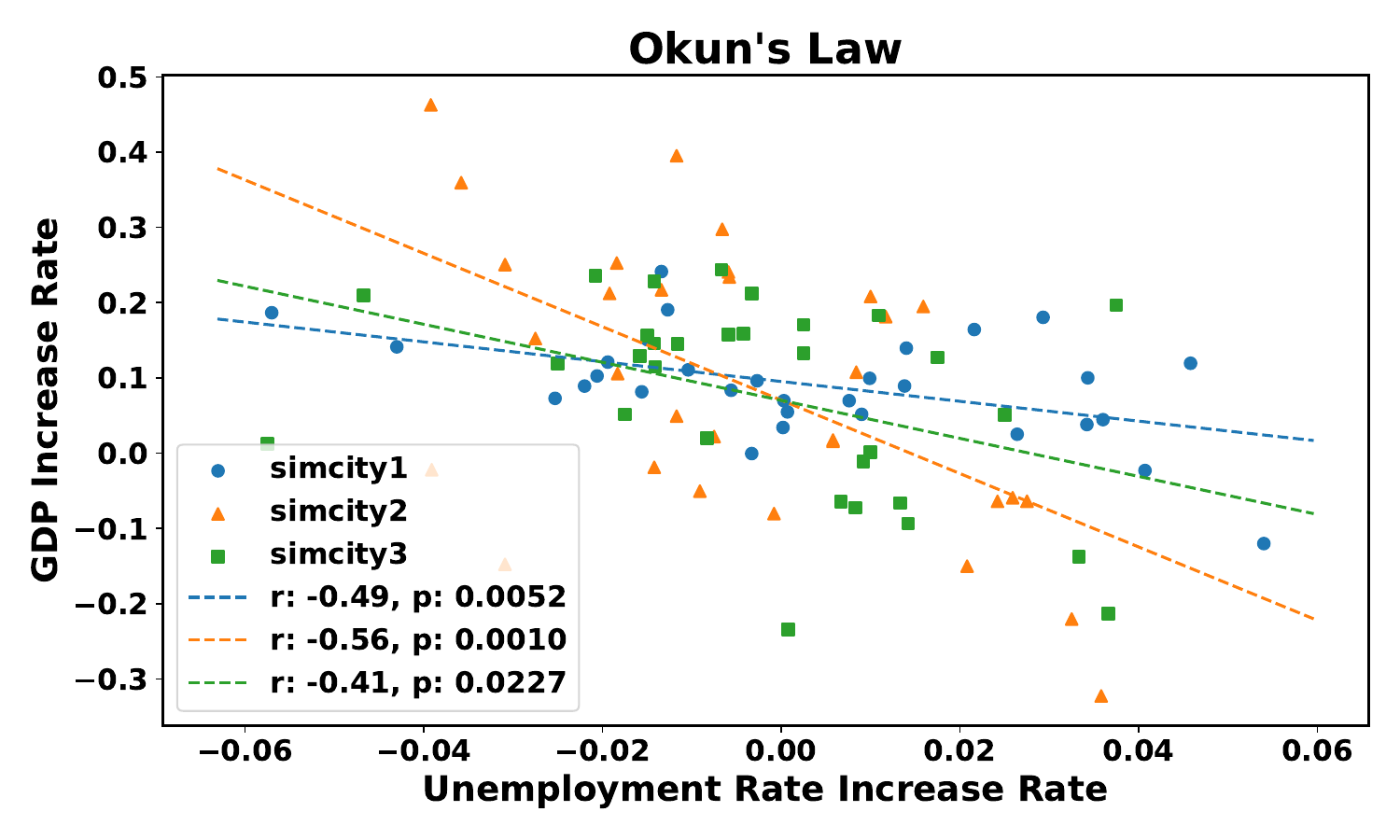}
\end{subfigure}
\begin{subfigure}[b]{0.31\textwidth}
\includegraphics[width=\textwidth]{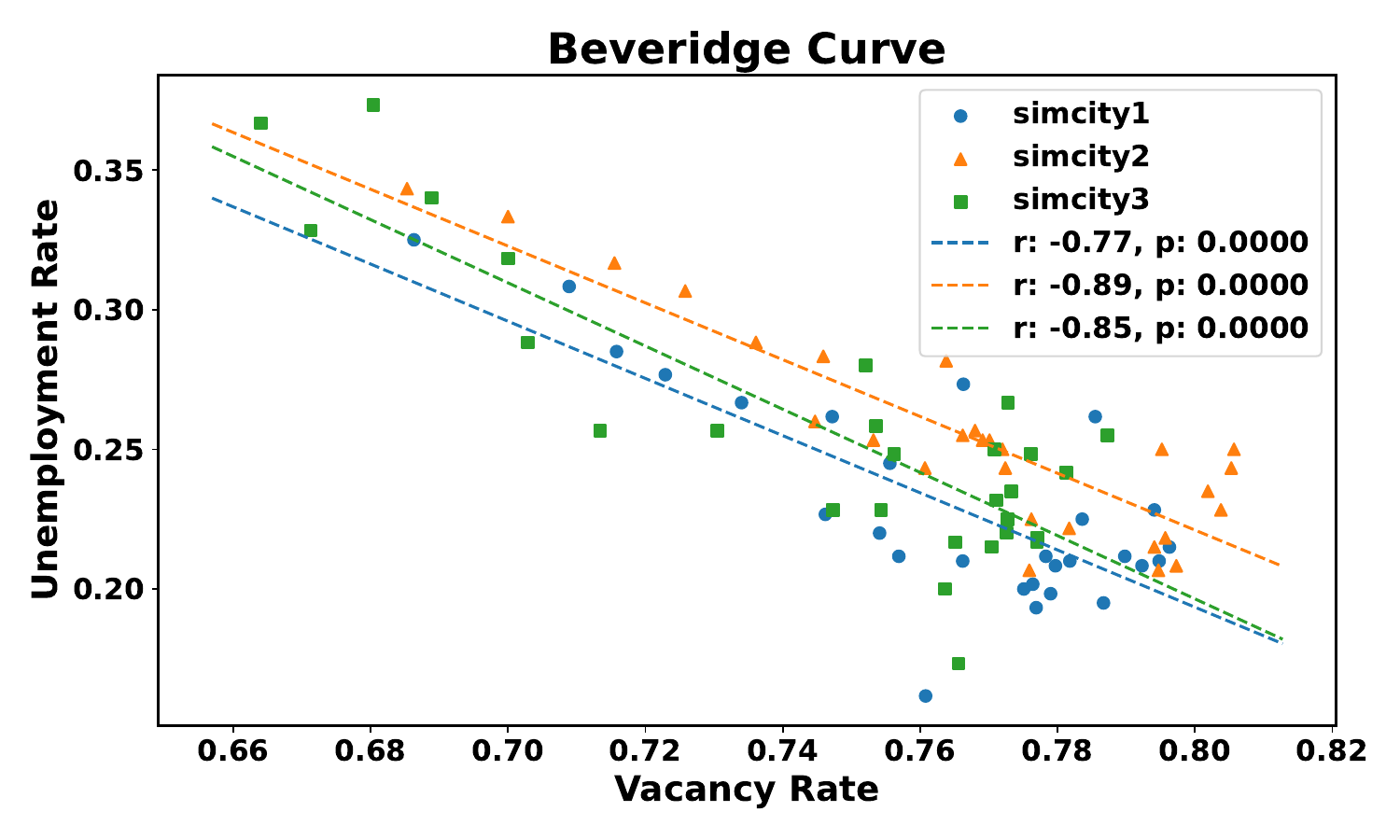}
\end{subfigure}
\end{minipage}
\caption{The results from different random seeds demonstrate that the observed regularity is robust.} \label{fig::stability-curves}
\end{figure*}

\begin{figure}[hbtp]
\vspace{-2mm}
\centering
    \begin{subfigure}[b]{0.49\linewidth}
        \includegraphics[width=\linewidth]{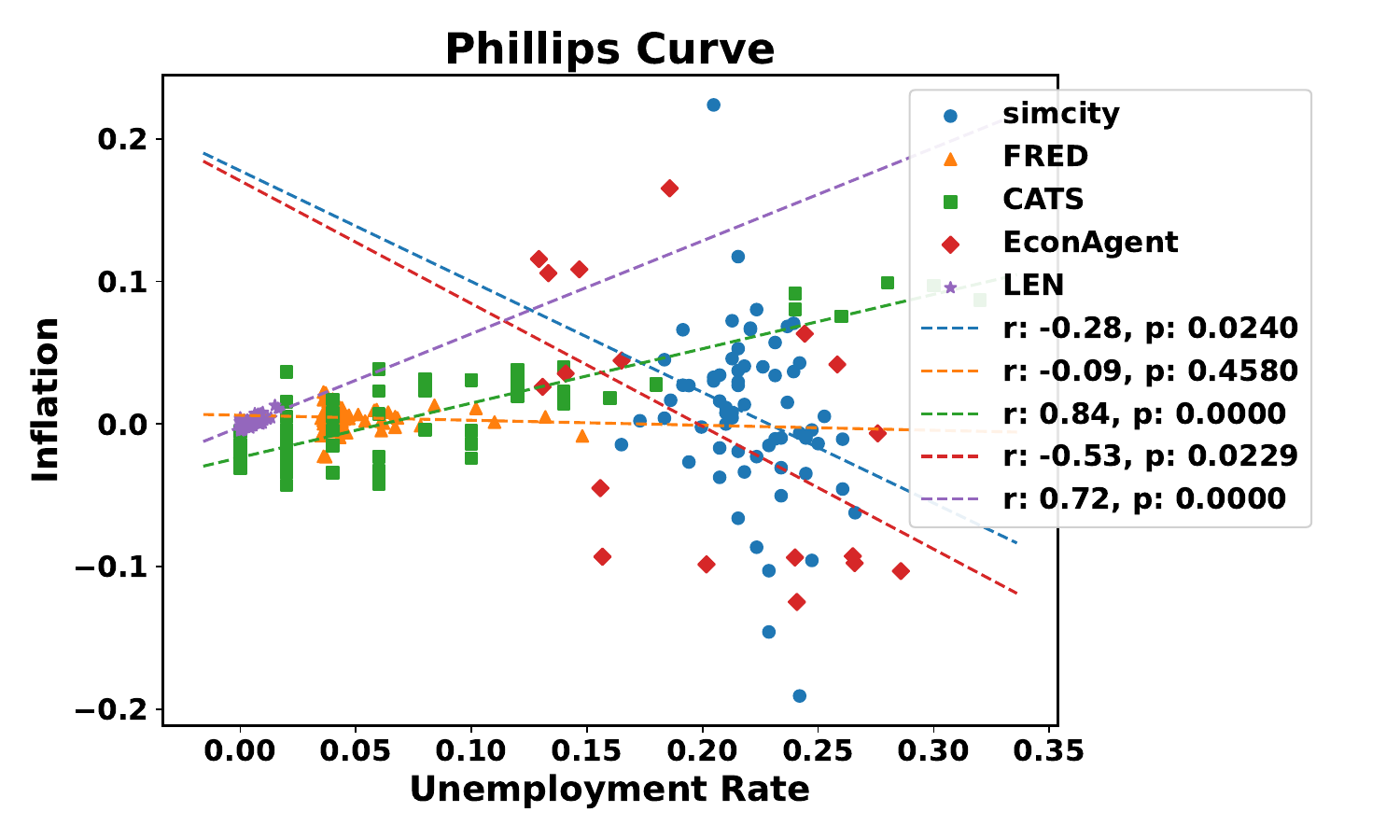}
    \end{subfigure}
    \vspace{-1mm}
    \begin{subfigure}[b]{0.49\linewidth}
        \includegraphics[width=\linewidth]{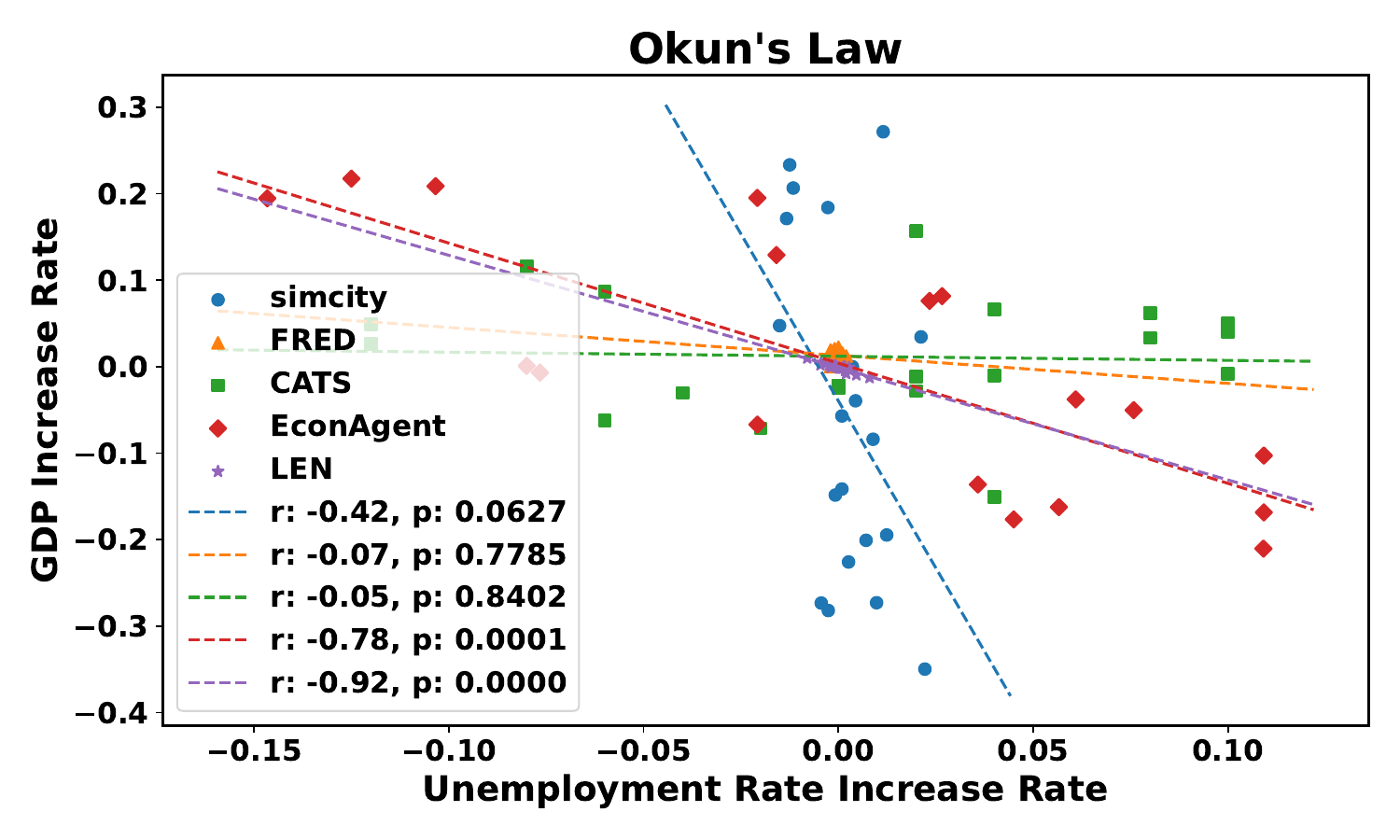}
    \end{subfigure}
    \caption{Macroeconomic patterns simulated using the \texttt{gpt-5} driven agents.}
    \label{fig::robustness-gpt5}
\end{figure}

Furthermore, we evaluate SimCity using \texttt{gpt-5} as the reasoning backbone. Figure~\ref{fig::robustness-gpt5} shows that the fundamental relationships remain consistent under a more advanced LLM.

\subsection{Price Impulse}\label{app:price_impulse}

\begin{figure}[hbtp]
\centering
\begin{minipage}{\textwidth}
\begin{subfigure}[b]{0.5\textwidth}
\includegraphics[width=\textwidth]{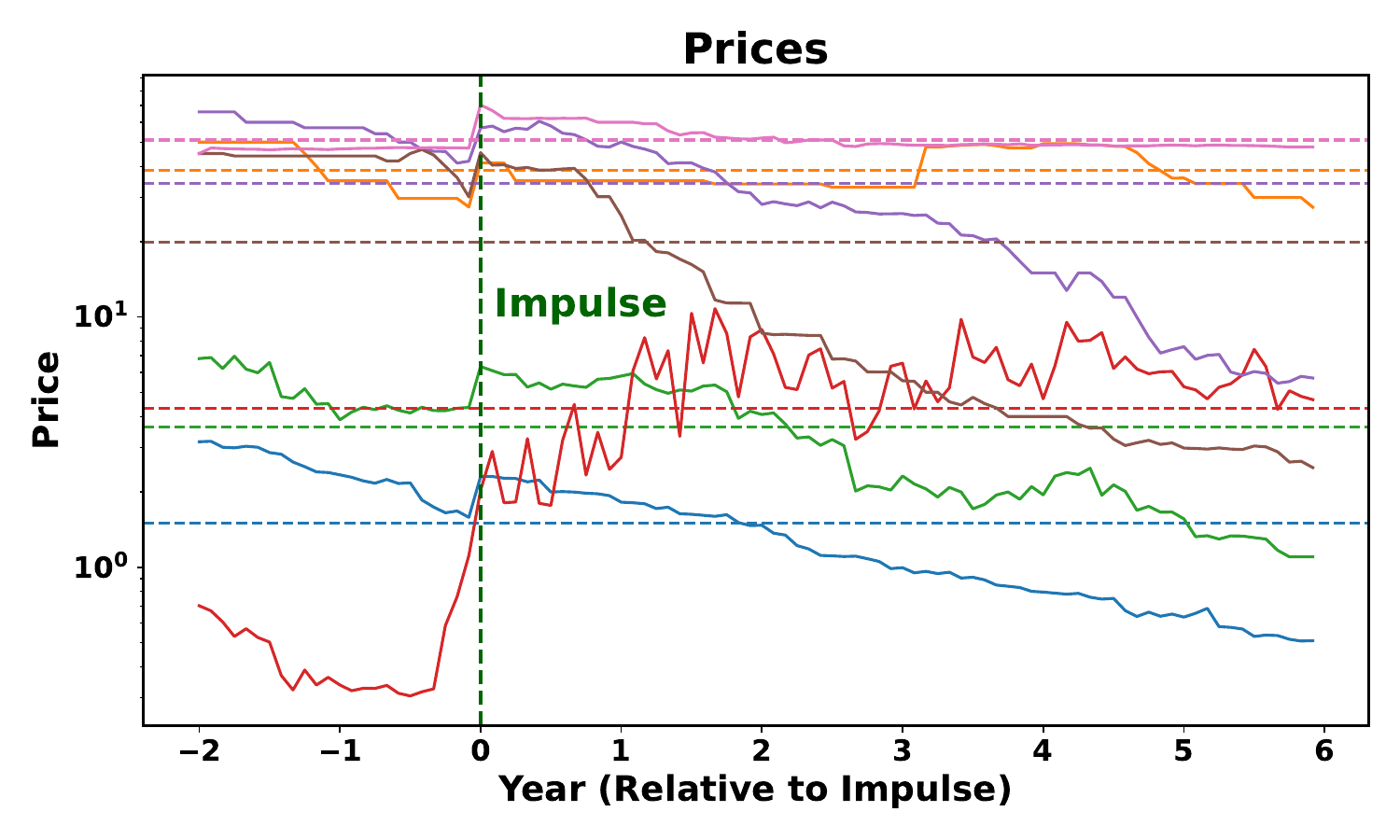}
\end{subfigure}
\begin{subfigure}[b]{0.4\textwidth}
\includegraphics[width=\textwidth]{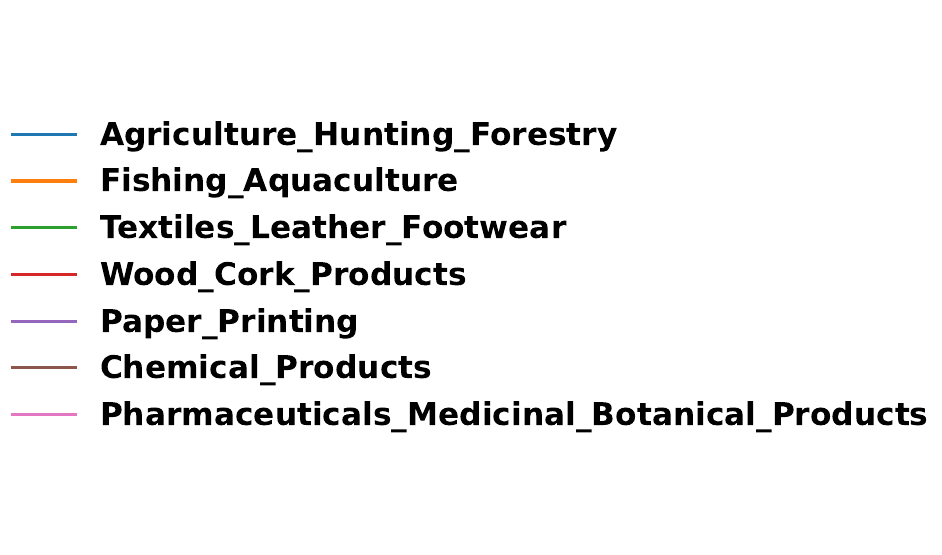}
\end{subfigure}
\end{minipage}
\caption{External impulse does not significantly affect long-run price trends.} \label{fig::price_up_impulse}
\end{figure}

The price-up impulse response, Figure~\ref{fig::price_up_impulse}, is shown here due to space limitations.

\subsection{Ablation Study: Visual Module}\label{app:visual_module_ablation}

To validate the contribution of the vision component, we compare the standard SimCity framework against a text-only variant. In this ablation setting, the establishment agent receives a list of building coordinates rather than a rendered map image to make location decisions.

\begin{figure}[h]
    \centering
    \begin{subfigure}[b]{0.45\textwidth}
        \centering
        \includegraphics[width=\textwidth]{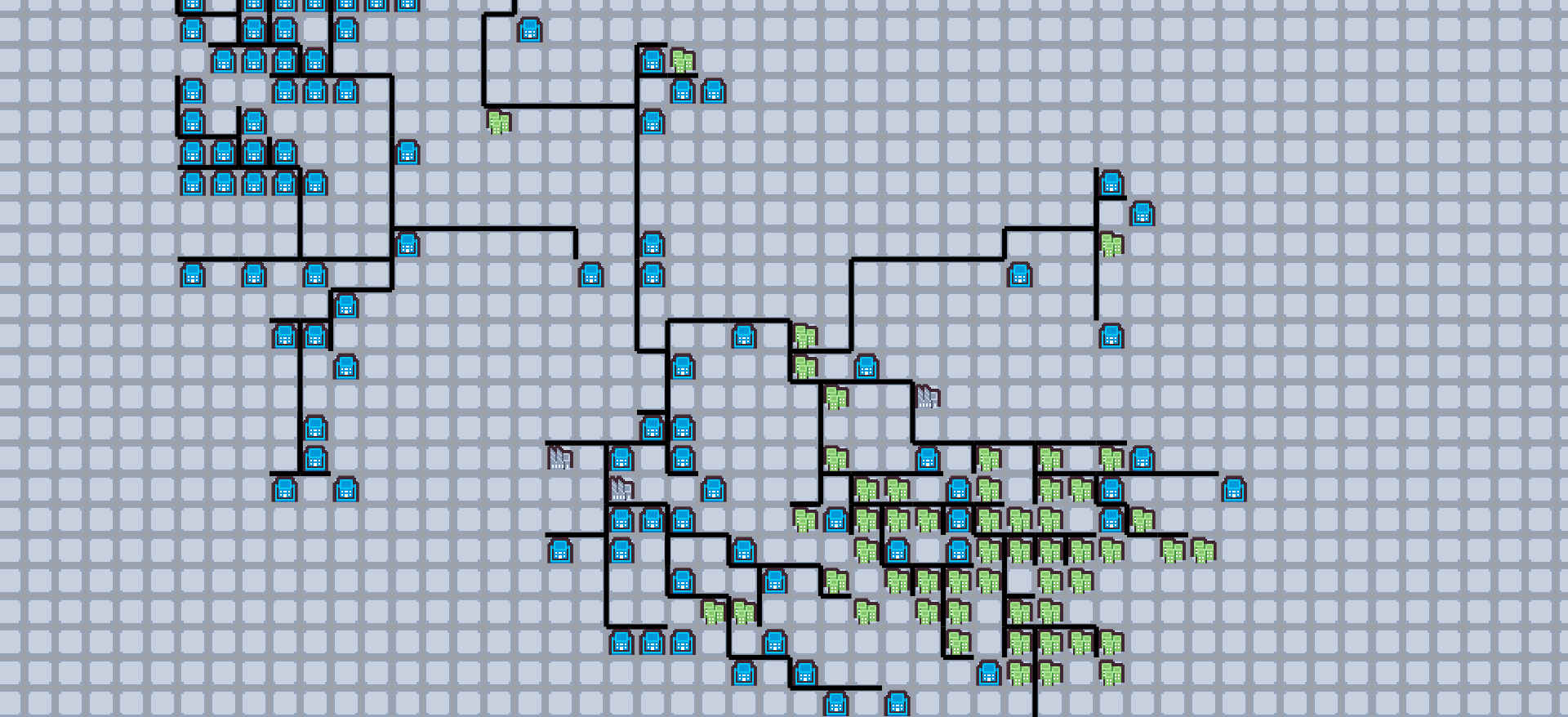}
        \caption{With VLM}
    \end{subfigure}
    \hfill
    \begin{subfigure}[b]{0.45\textwidth}
        \centering
        \includegraphics[width=\textwidth]{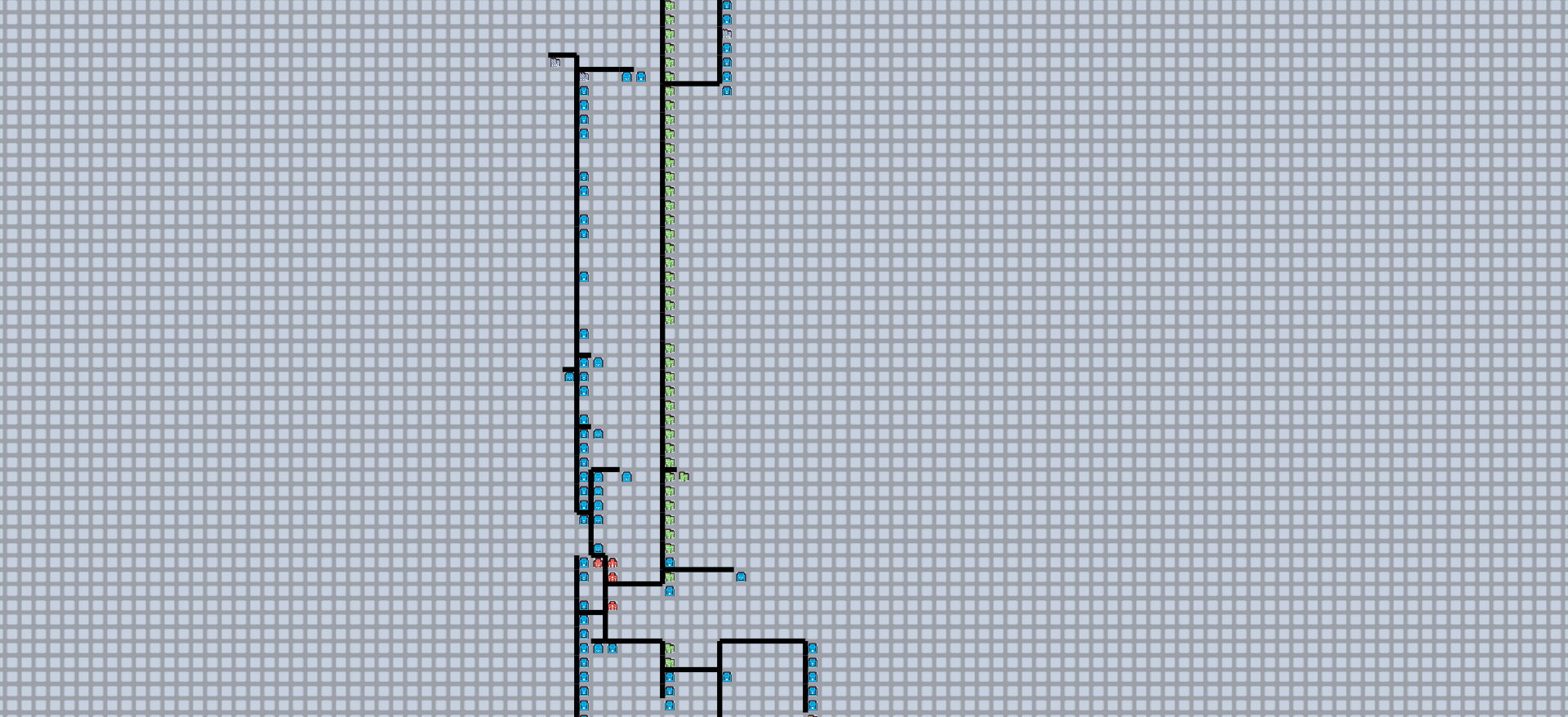}
        \caption{Text Description Only}
    \end{subfigure}
    \caption{Impact of the Visual Module on urban layout. The full model produces a realistic, clustered city center, whereas removing the visual input results in an unrealistic linear topology.} 
    \label{fig::visual_module_ablation}
\end{figure}

As illustrated in Figure~\ref{fig::visual_module_ablation}, the absence of the visual module significantly degrades the spatial logic of urban planning. While the VLM-based approach (Figure~\ref{fig::visual_module_ablation}a) fosters a natural, concentric city structure with a distinct center, the text-only approach (Figure~\ref{fig::visual_module_ablation}b) fails to interpret spatial density effectively. Consequently, the agents form an artificial linear arrangement of buildings ("two lines"), confirming that the visual module is crucial for interpreting complex spatial information that is difficult to convey through text alone.

\subsection{Extra Example for City Expansion}\label{app:extra_example_city_expansion}

\begin{figure}
    \centering
    \includegraphics[width=\linewidth]{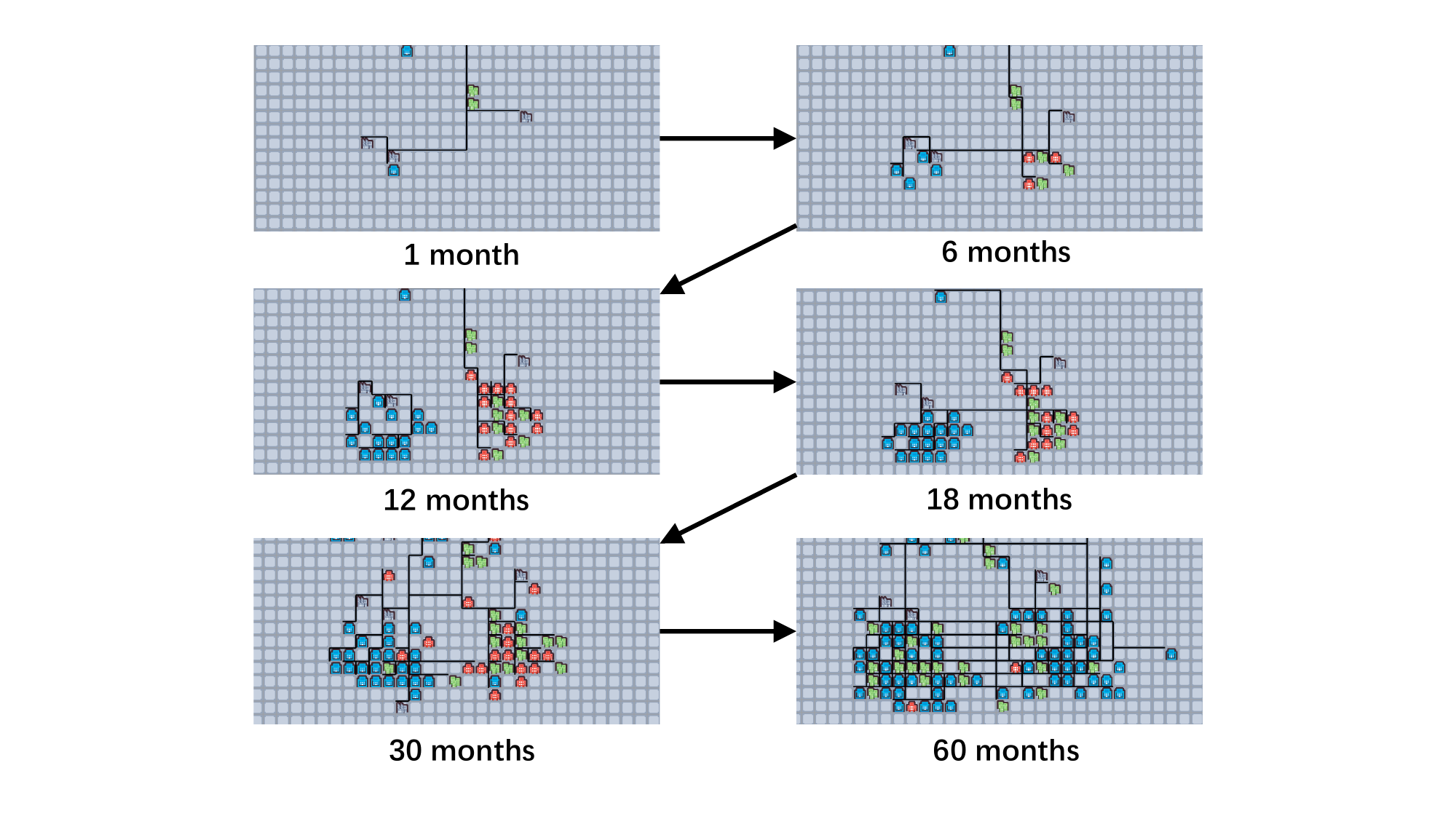}
    \caption{Another example of city dynamics during the move-in phase.}
    \label{fig::extra_example_city_expansion}
\end{figure}

An extra example for city expansion, Figure~\ref{fig::extra_example_city_expansion}, is shown here.

\section{Agents Details}\label{app:agents_definition}

\subsection{Notations}

We define the following notations:
\begin{itemize}
    \item The goods set, $\mathcal{G}$, where $|\mathcal{G}|$ denotes the number of heterogeneous goods present in our simulation. Certain subsets of $\mathcal{G}$ have a specific meaning, such as $\mathcal{G}_E$ for essential goods, $\mathcal{G}_T$ for transportation goods, and $\mathcal{G}_D$ for durable capital.
    \item Each step in our simulation represents a month. For clarity, time subscripts are suppressed unless needed for exposition. For example, the labor choice at time $t$ for household $i$ is written as \(l_i\) instead of \(l_{i,t}\).
\end{itemize}

\subsection{Households}\label{app:households_detail}
A household \( i \) is characterized by the following attributes:
\begin{itemize}
  \item \textbf{Cash Holdings \( m_i \):} The amount of money held by the household. 
  \item \textbf{Labor Participation:} An indicator \( l_i \in \{0, 1\} \), where 1 means the household is employed to work in a position denoted by \( \mathrm{position}_j \) and receives a salary \( S_j \), and 0 means the household is unemployed. Note that we are modeling the extensive margin of labor, rather than of the intensive margin (hours worked).
  
  \item \textbf{Innate Skills:} Each household is endowed with heterogeneous skills.  Please see Table~\ref{table:generated_skills_4col} for the full list of skills. For each skill \( j \), the innate skill level \( s_{i,j} \sim U(s_{\min}, s_{\max}) \)
  means that when the household works in a job requiring skill \( j \), it supplies \( l_i \times s_{i,j} \) effective units of labor. Additionally, the probability that a household is matched with a vacancy requiring skill \(j\) is an increasing function of \(s_{i,j}\). This is in line with the directed search literature in economics \citep{wright2021directed}.
  
  \item \textbf{Heterogeneous Consumption Preferences:} The household has needs denoted by \( n_i \). For each good \( g \in \mathcal{G} \), \( n_{i,g} \) represents the desired number of units of good \( g \) to be purchased. \(n_i\) may split to essential needs \(n_{e, i}\) which is fixed denoting immutable demands and additional needs \(n_{a, i}\). The modeling of essential goods has a long history in economics. See, for example, the Stone-Geary Utility function \citep{Stone1954LinearExpenditure}.

  \item \textbf{Housing:} Each household resides in a house \( \mathcal{H}_i \). Consequently, the household must pay rent \( r_i \) to the owner of the property\footnote{In this model, we assume that everyone rents their housing. In reality, if a household owns its home, one can think of it as renting a property in which it holds full ownership.}.
\end{itemize}

\subsection{Firms}\label{app:firm_detail}

Each firm \( i \) is characterized by the following attributes:
\begin{itemize}
    \item \textbf{Cash Holdings \( m_i \):} The monetary resources currently available to the firm.
    \item \textbf{Output Good:} Each firm specifies a specific type of good \( g_i \in \mathcal{ G} \).
    \item \textbf{Shareholders:} A collection of households and/or government entities, each holding a share denoted by \( \textbf{share}_{i, j} \).
    \item \textbf{Dividend Rate \( d_i \):} A parameter \( d_i \in [0, 1] \) representing the fraction of the firm's profits that are distributed as dividends to shareholders. Specifically, a shareholder holding \( \textbf{share}_{i, j} \) receives a dividend computed as:
    \[
        D_j = \frac{\textbf{share}_{i, j} \, d_i \, m_i}{\sum_k\textbf{share}_{i,k}}.
    \]
    % \item \textbf{Efficiency \( e_i \):} A metric \( e_i \in [0, 1] \) indicating the firm's operational efficiency. A value below 1 signifies that the firm is operating below its full capacity (i.e., “idling”) to reduce production output and material costs.
    % \item \textbf{Price Settings \( \mathcal{P}_i \):} The set of prices at which the firm offers its products.
    % Where \(\textbf{share}_{i}\) denotes the sum of the shares of all shareholders in firm \(i\).
    \item \textbf{Job Positions \( \mathcal{J}_i \):} A list of job vacancies within the firm. Each position \( j \) requires a specific skill and offers a salary \( S_{i,j} \).
    \item \textbf{Durable Capital \( \mathcal{K}_i \):} The stock of durable capital assets that the firm has invested in to support production.
    \item \textbf{Estimated Value \( \mathcal{V}_i \):} The firm's estimated value, calculated as:
    \[
        \mathcal{V}_i = \frac{\sum_{j = t - 12}^{t - 1}\textbf{profit}_{j}}{12\cdot \max(\mathcal{I}_{d, t}, \epsilon_d)} + m_i - L_i + \mathcal{K}_i\cdot 
    \]
    where \(\mathcal{I}_{d, t}\) denotes the GDP deflator, \(\epsilon_d = 0.02\) is a lower bound introduced to prevent division by zero and \(\textbf{profit}_{j}\) denotes the net profit in month \(j\), defined as the sales revenue minus the production expenditures.
    \item \textbf{Loans}: The amount of loans the firm owes to the financial system.
\end{itemize}

If a firm experiences zero income for a year and its assets are insufficient to cover its overdue debts, the firm is declared bankrupt. In that case, all employees are terminated, and the land it occupies is released.

A rentable apartment (residential building) is also formally modeled as a firm, though its decision-making is limited to adjusting its rental price based on conditions in the housing market.

\subsection{Government}\label{app:government_detail}

The government records a series of indicators:
\begin{itemize}
    \item \(\mathbf{nominalGDP}_t\): the sum of consumption, investment, and government spending, and \(\mathbf{realGDP}_t\), computed things above using the base year's prices.
    \item \(\mathbf{eq}_t\), \(\mathbf{giniW}_t\), \(\mathbf{giniI}_t\): measures of equality, wealth inequality (Gini coefficient), and income inequality (Gini coefficient), defined as follows. Suppose there are \(n\) households, and let \(\{x_{(i)}\}_{i=1}^n\) denote the sorted (in increasing order) wealth (or income) of households. Then,
\begin{align}
\textbf{giniW}_t &= \frac{\sum_i \text{sorted\_wealth}_i \cdot 2i}{n\cdot \text{sum\_wealth} - \frac{n+1}{n}}, \\
\textbf{giniI}_t &= \frac{\sum_i \text{sorted\_income}_i \cdot 2i}{n\cdot \text{sum\_income} - \frac{n+1}{n}}, \\
\textbf{eq}_t &= 1 - \frac{n}{n - 1}\textbf{giniI}_t.
\end{align}
    \item Unemployment rate \(u_t\), along with broad monetary aggregates \(\text{M}0\) and \(\text{M}1\)\footnote{\(\text{M}2\) is not included since fixed deposits have not been defined.}.
    % \item Happiness measure \(\text{happiness}_t\).
    \item Goods production \(Y_t\), and average production \(\bar{Y} = \frac{\sum_t Y_t}{T}\).
    \item Inflation: including wage inflation \(\mathcal{I}_w\) and GDP inflation \(\mathcal{I}_d\), each defined by:
\begin{align}
\mathcal{I}_{w, t} &= \frac{\sum_{\text{household}} S_{i, t}}{\sum_{\text{household}} S_{_i, t-1}} - 1,\\
\mathbf{GDPdeflator}_t &= \frac{\mathbf{nominalGDP}_t}{\mathbf{realGDP}_t}, \\
\mathcal{I}_{d, t} &= \frac{\mathbf{GDPdeflator}_t}{\mathbf{GDPdeflator}_{t - 1}} - 1.
\end{align}
\end{itemize}

\subsection{Central Bank and Financial System}\label{app:cb}
We implement a Taylor rule with interest rate smoothing in the policy behavior of our simulated central bank agent. Without smoothing, simulated interest rates exhibit implausibly high volatility in response to small shocks. 

Interest rate smoothing refers to the empirical regularity that central banks adjust policy rates gradually over time, rather than immediately to the level implied by contemporaneous macroeconomic conditions. This behavior is often interpreted as reflecting forward-looking policy preferences, model uncertainty, or an aversion to financial market instability. In practice, this means that the actual interest rate set by the central bank is 
\begin{equation}
    r_{t}=\rho r_{t-1} + (1-\rho)\hat{r}_t,
\end{equation}
where \(\hat{r}\) is the policy rate computed from the unmodified Taylor rule~\ref{eq:taylor_rule}, \(r_{t-1}\) the interest rate set last period, and \(\rho\) is the smoothing factor.

Empirically, interest rate smoothing is a well-documented phenomenon. For example, \citet{10.1257/mac.4.4.126} estimates a Taylor rule with smoothing and finds that the smoothing coefficient typically ranges between 0.7 and 0.9 in developed economies. We choose \(\rho=0.8\) as our parameterization.

\section{Framework Details}

\subsection{Environment Setup}\label{app:environment_setup}

There are 44 types of goods in SimCity. Each of them represents an economic industry in the real world as categorized by the Organization for Economic Co-operation and Development (OECD). 

Initial prices of all goods are set at 50. However, the specific choice of initial prices does not matter for our simulation. We experimented with various initial prices, but in all cases, as we mention in~\ref{sec::price_impulse}, prices reach steady-state in a small number of steps. Similarly, wages of positions and rents of residential buildings are set arbitrarily. 

\subsection{Stages within Each Step}\label{app:stages}

As aforementioned, each simulation step represents one month, and each step includes the following four stages:

\begin{itemize}
    \item \textbf{Production and Trading Stage:} Firms produce goods, after which households and firms purchase goods for consumption and materials as planned.
    \item \textbf{Taxation and Dividend Stage:} Firms pay dividends, while the government collects taxes and disburses welfare.
    \item \textbf{Metabolic Stage:} New companies are established through equity financing from the investment pool, bankrupt companies are removed, and the population grows according to set rules.
    \item \textbf{Revision Stage:} Households, firms, the government, and the central bank agents review their situations and decide on their actions. Only the revision stage involves LLM agents.
\end{itemize}

\subsection{Render Module}\label{app:render_module}

We utilize \texttt{flask-socketio} to build a web server and \texttt{Vue.js} to build a website.

The assets are from the open-source \texttt{tiny-battle} package created by Kenney\footnote{\url{https://www.kenney.nl/}}.
% Render Module is 

\subsection{Profile Setup}\label{app:profile_setup}
Age distribution is from the Demographic and Housing Characteristics (DHC) table from the U.S. Census Bureau\footnote{ \url{https://data.census.gov/table?q=PCT12&d=DEC+Demographic+and+Housing+Characteristics}}.

To assign realistic initial cash holdings to households, we estimate the income distribution using U.S. microdata. We use the 2023 American Community Survey (ACS) IPUMS microdata, which contains detailed household-level income information. We assume a lognormal distribution for household income, a standard approximation in the economics literature due to its simplicity and its ability to capture the right-skewed nature of income data.\footnote{While some heavier-tailed distributions (e.g., Pareto) better approximate top incomes, our simulation does not focus on the super-rich. Hence, the lognormal distribution is a reasonable choice. See \citet{https://doi.org/10.1111/joes.12435} for a recent summary of related work.} Formally, the initial cash holding is drawn from
\begin{equation}
    \ln m_{i, 0} \sim \mathcal{N}(\mu, \sigma^2) 
\end{equation}
where \(\mu = 11.1496, \sigma^2=1.1455\), which are estimated via maximum likelihood estimation (MLE) with the 2023 ACS microdata.

\subsection{Synthesis}\label{app:synthesis_firm}
We assume that each category of goods is produced by a representative type of firm. Therefore, we generate a firm template for each category, which specifies the firm's name, the skill requirements for its positions, and an input-output "recipe" derived from the Input-Output Tables (IOTable) provided by the OECD.\footnote{\url{https://www.oecd.org/en/data/datasets/input-output-tables.html}}

We process the IOTable to determine the inputs for each firm. For each category of goods, we normalize the input requirements from all other categories needed to produce one unit of output. We then select the top input categories that cumulatively account for more than 75\% of the total input value, defining these as the necessary inputs for the corresponding firm's production process.

Then, we use the prompt introduced in Listing~\ref{lst:synthesis-prompt-factory-maker} to call a LLM to generate the processed input-output into a firm template.

Observing that the positions required skills in the generated firm templates contain many semantically similar skills, we use the prompt introduced in Listing~\ref{lst:synthesis-prompt-merge-skill} to call an LLM to merge these similar skills. Skills after merge are shown in Table~\ref{table:generated_skills_4col}.

\begin{lstlisting}[label=lst:synthesis-prompt-factory-maker, caption=Prompt for firm template generation.]
I am currently developing a game. One of the tasks I need to do now is: Given a recipe, please help me design a building to produce this recipe.
First, I will give you a recipe, which is composed of some raw materials. Then, you need to design a suitable building to produce it. Noticing that the employees' core skill should be as abstract as possible.

``type`` must either "BusinessBuilding" or "ServiceBuilding". If you consider a type of product is mainly provided by government, you should set it as ServiceBuilding, else BusinessBuilding.
Here is an example:


```
### Recipe ###

{
  "input": {
    "RawFood": 1000
  },
  "output": {
    "Food": 1000
  }
}

Your Response:

  {
    "template_name": "Restaurant",
    "width": 1,
    "height": 1,
    "description": "A restaurant that uses raw food to serve food.",
    "type": "BusinessBuilding",
    "building_cost": 500000,
    "provide_radius": 10,
    "employees": [
      {
        "position_type": "Manager",
        "salary": 3000.0,
        "core_skills": [
          "Management"
        ],
        "importance": 1.0
      },
      {
        "position_type": "Cook",
        "salary": 1000.0,
        "core_skills": [
          "Cooking"
        ],
        "importance": 0.5
      },
      {
        "position_type": "Waiter",
        "salary": 1000.0,
        "core_skills": [
          "Service"
        ],
        "importance": 0.5
      },
      {
        "position_type": "Waiter",
        "salary": 1000.0,
        "core_skills": [
          "Service"
        ],
        "importance": 0.5
      }
    ],
    "recipes": [
      {
        "input": {
          "RawFood": 1000
        },
        "output": {
          "Food": 1000
        }
      }
    ]
  }

```

Now the recipe I provide will listed follow "### Recipe ###"
You should respond with a factory design in the same format as the examples above.
\end{lstlisting}

\begin{lstlisting}[label=lst:synthesis-prompt-merge-skill, caption=Prompt for merging skills]
```
[Skills]
```
The content above is a list of several skills from my economic simulation environment, SimCity. A job posting might require several of these skills. I now feel that there are too many of them and they need to be simplified. Please merge the skills that have the same meaning into a single skill (note: only merge two skills if and only if they are very similar; for example, Geology and Mining should not be merged because they are different specialties), and provide the merge relationships: a JSON object { "A": "B" } indicates that skill A is essentially skill B.

If several skills are to be merged into one category, choose the most representative one as B. Then, return this JSON object.
\end{lstlisting}

\begin{table*}[h!]
\centering
\small
\begin{tabular}{llll}
\hline
Management & Quality Control & Operations Management & Logistics Management \\
Physical Labor & Aquaculture & Environmental Science & Engineering \\
Monitoring & Maintenance & Geology & Mining \\
Chemistry & Surveying & Technical Skills & Food Science \\
Machinery Operation & Safety Management & Supply Chain Management & Pharmaceutical Science \\
Laboratory Skills & Metalworking & Assembly & Energy Management \\
Water Management & Equipment Maintenance & Project Management & Design \\
Building & Sales & Customer Service & Vehicle Maintenance \\
Data Analysis & Regulations & Equipment Operation & Driving \\
Transportation & Communication & Culinary & Cleaning \\
Media Production & Equipment Handling & Logistics & Information Technology \\
Finance & Technical Support & Real Estate & Insurance \\
Market Analysis & Office Management & Administrative Support & Building Maintenance \\
Research & Consulting & Facility Maintenance & Basic Repairs \\
Support & Teaching & Marketing & Legal Knowledge \\
Human Resources & Sanitation & & \\
\hline
\end{tabular}
\caption{List of skills.}
\label{table:generated_skills_4col}
\end{table*}

\newpage
\section{Prompting and Examples}

\definecolor{promptbg}{rgb}{0.97,0.97,0.97}
\definecolor{promptborder}{rgb}{0.7,0.7,0.7}
\definecolor{rea}{RGB}{0, 158, 115}    
\definecolor{ac}{RGB}{213, 94, 0}  
\definecolor{act}{RGB}{0, 114, 178}
\definecolor{res}{RGB}{204,121,167}
\definecolor{success}{RGB}{0,128,0}

\captionsetup{hypcap=false}

\tcbset{
  promptstyle/.style={
    enhanced,
    colback=promptbg,
    colframe=promptborder,
    fonttitle=\bfseries,
    coltitle=black,
    sharp corners,
    boxrule=0.4pt,
    left=6pt,
    right=6pt,
    top=4pt,
    bottom=4pt,
    boxsep=2pt
  }
}

% \lstset{
% basicstyle=\rmfamily\normalsize,    
%   breaklines=true,               
%   breakatwhitespace=true,        
%   columns=fullflexible,
%   breakindent=0pt,
%    literate=
%     {'}{\textquotesingle}1
%     {`}{\textasciigrave}1,   
% }

\lstdefinestyle{logcode}{
    language=Python,
    basicstyle=\ttfamily\small, 
    numberstyle=\tiny\color{gray}, 
    keywordstyle=\color{blue}, 
    commentstyle=\color{green!50!black}, 
    stringstyle=\color{orange},
    breaklines=true,
    frame=single,
    showstringspaces=false,
    tabsize=4
}

\lstdefinestyle{reasoning}{
    basicstyle=\rmfamily\normalsize\color{rea},
    aboveskip=0pt,
    belowskip=0pt
}

\lstdefinestyle{action}{
    basicstyle=\rmfamily\normalsize\color{ac},
    numbers=none,
    identifierstyle=\color{ac},
    stringstyle=\color{ac},
    commentstyle=\color{ac},
    literate={
        {`}{{\color{ac}`}}1
        {'}{{\color{ac}'}}1
        {/}{{\color{ac}/}}1
        {.}{{\color{ac}.}}1
    },
    aboveskip=0pt,
    belowskip=0pt
}

\lstdefinestyle{action_content}{
    basicstyle=\rmfamily\normalsize\color{act},
    numbers=none,
    identifierstyle=\color{act},
    stringstyle=\color{act},
    commentstyle=\color{act},
    literate={
        {`}{{\color{act}`}}1
        {'}{{\color{act}'}}1
        {/}{{\color{act}/}}1
        {.}{{\color{act}.}}1
    },
    aboveskip=0pt,
    belowskip=0pt
}

\lstdefinestyle{response}{
  basicstyle=\rmfamily\normalsize\color{res},
    aboveskip=0pt,
    belowskip=0pt
}
\begin{tcolorbox}[promptstyle, breakable, title=Structure of Prompt]

% \textcolor{red}{\textbf{Starting iteration 26/100}}\\
\textbf{System Prompt:}
\begin{lstlisting}[style=response]
You are a citizen of SimCity and you are taking action improve your life. Basically speaking, your goal is to be consuming a greater variety of products, acquiring more money and assets, and having a stable job and residence.
There are some information about you and the city which may affect your life...
\end{lstlisting}

\textbf{User Prompt:}
\begin{lstlisting}[style=reasoning]
### Profile 
Your name is [name], and you are [age] years old...... You current have [money] money. You are working at [place] as [position] with salary [salary]...
### Report
...During the past 12 months, your average income is [average_income]. This month, your income is [total_income] , which consists of the following parts:
  - Salary: ...
  - Benefit: ...
...
This month, your outcome is [total_outcome]
...
### Observation
Here are opening positions in the city:
  -[position_id]: ...
...
The social average return on investment is [roi], and the interest rate of the bank is [interest_rate].
\end{lstlisting}
\captionof{figure}{Structure of prompts.}
\label{fig::citizen_prompt_example}
\end{tcolorbox}

% \begin{wrapfigure}{r}{0.6\linewidth}
% % \centering
%     \includegraphics[trim=0 0 0 0,clip,width=\linewidth]{Fig/prompt_citizen.pdf}
%     \caption{Structure of prompt}
%     \label{fig::citizen_prompt_example}
% \end{wrapfigure}

\subsection{Prompting Structure}\label{app:prompting}
We aim to leverage the common-sense capabilities of large language models to act as human-like, heterogeneous agents. Figure \ref{fig::citizen_prompt_example} provides the structure of the prompt for a household agent.

As mentioned in Section~\ref{sec::modeling}, the Agent will receive the necessary information to make its decisions. The Profile includes some basic personal information about the Agent, such as current income, age, skills, job, and residence. The Report includes some of the Agent's recent experiences and information. The Observation presents information about the overall environment that the Agent should be aware of, such as vacant positions, the unemployment rate, prices of goods, and the Return on Investment (ROI).

Prompts for other agents are structured in a similar manner. Detailed examples are provided in  Appendix~\ref{app:prompt_examples}.

% \begin{figure}
%     \centering
%     \includegraphics[width=0.6\linewidth]{Fig/prompt_citizen.pdf}
%     \caption{Structure of dialogue of an agent.}
%     \label{fig:placeholder}
% \end{figure}

\subsection{Function Calling}\label{app:function_calling}
The model interacts with the environment by means of Function Calling. The framework loads all the operations that the Agent can execute and appends formatted function names along with their descriptions to the prompt. The LLMs will return the actions to be taken and their parameters in JSON format. The framework will execute after a verification. 

\subsection{Full Examples}\label{app:prompt_examples}

Full examples of all types of agents can be found at Table~\ref{table:prompt_examples}.

\begin{table}[t]
\small
\setlength{\tabcolsep}{2pt}
\centering
\begin{tabular}{lccc}
\multicolumn{1}{c}{\bf Agent Type}  &\multicolumn{1}{c}{\bf System Prompt}
&\multicolumn{1}{c}{\bf User Prompt}
&\multicolumn{1}{c}{\bf Response}
\\ \hline \\
{Household} & {Listing~\ref{lst:household-agent-system-prompt}} & {Listing~\ref{lst:household-agent-user-prompt}} & {Listing~\ref{lst:household-agent-response-example}} \\
{Government} & {Listing~\ref{lst:government-agent-system-prompt}} & {Listing~\ref{lst:government-agent-user-prompt}} & {Listing~\ref{lst:government-agent-response-example}} \\
{Investment Pool} & {Listing~\ref{lst:investment-pool-agent-system-prompt}} & {Listing~\ref{lst:investment-pool-agent-user-prompt}} & {Listing~\ref{lst:investment-pool-agent-response-example}} \\
{Firm(Productive Building)} & {Listing~\ref{lst:productive-firm-agent-system-prompt}} & {Listing~\ref{lst:productive-firm-agent-user-prompt}} & {Listing~\ref{lst:productive-firm-agent-response-example}} \\
{Firm(Residential Building)} & {Listing~\ref{lst:residential-firm-agent-system-prompt}} & {Listing~\ref{lst:residential-firm-agent-user-prompt}} & {Listing~\ref{lst:residential-firm-agent-response-example}} \\
 \\ \hline \\
\end{tabular}
\caption{Example Listings}
\label{table:prompt_examples}
\end{table}

% \newpage
% --- Household Agent ---
\begin{lstlisting}[label=lst:household-agent-system-prompt, caption=System prompt example of household agent.]
You are a citizen of SimCity and you are taking action improve your life. Basically speaking, your goal is to be consuming a greater variety of products, acquiring more money and assets, and having a stable job and residence.
There are some information about you and the city which may affect your life. The information including four parts: "### Profile", "### Report", "### Observation", and "### Actions".
You will be provided your personality traits and skills, and some personal information, which will be listed after "### Profile".
You have also provided your last month's living status, which will be listed after "### Report".'
Any information about the city which may affect your life will be listed after "### Observation".
Based on this information, you may take some actions to improve your life. These actions you may take will be listed after "### Actions".

You should only respond in JSON format as described below:
{
    "reasoning": "reasoning"
    [
        "action1(para1, para1, ...)",
        "action2(para1, para1, ...)",
        ...
    ]
}

Ensure the response can be parsed by Python `json.loads`, e.g.: no trailing commas, no single quotes, etc.
\end{lstlisting}

\begin{lstlisting}[label=lst:household-agent-user-prompt, caption=User prompt example of household agent.]
### Profile

Your name is David, and you are 54 years old. Your ID is 2000161.You current have 1878.741 money in your account.
You are a Disorganized,Unadventurous woman
You are currently unemployed.

You are living in Residential Building at (25, 15) with rent 450. 

You plan to spend a percentage of your income on the following items, at a total of 1.11% of your income:
  - Agriculture_Hunting_Forestry: 0.0
  - Fishing_Aquaculture: 0.01
[... and other needs.]
  - Household_Employer_Activities: 0.0
  - Motor_Vehicles_Trailers_SemiTrailers: 0.0

Bank Account David:
You have no savings
You have no loans.
Demand deposit interest rate is 0.6108%, and loan rate is 2.6108%.
Your loanable amount is 243757.48131444445.


### Report

This month, your income is 2878.741, which consists of the following parts:
  - Salary: 1000.000(34.7374%)
  - Benefit: 1000.000(34.7374%)
  - Profit: 0.000(0.0000%)
  - Welfare: 878.741(30.5252%)

This month, your outcome is 1442.924, which consists of the following parts:
  - Consumption: 1017.924(70.5459%)
    - Agriculture_Hunting_Forestry: 297.363(29.2127%)
    - Food_Beverages: 720.561(70.7873%)
  - Rent: 425.000(29.4541%)


### Observation

Here are opening positions in the city:
  - position_id: 2003486, Production Manager requires skills Management, Quality Control, with salary 600.000
[... and other available positions.]
  - position_id: 2000253, Plant Manager requires skills Management, Quality Control, with salary 5000.000
You expect the work should cover your spending of 1442.924.

Here are available residential buildings in the city:
  - residential id: 200089 at (4, 3) with rent 450
[... and other residential buildings' position and rent]
  - residential id: 200366 at (18, 19) with rent 450

The social average return on investment is 1.3949%, and the interest rate of the bank is 0.6108%.
If roi is higher than the interest rate of the bank, you can invest your money in the investment market to get a higher return on investment and you may borrow money from the bank to invest if you need.
Otherwise, you may simply save your money in the bank.

The prices of goods on the market are as follows:
  - Agriculture_Hunting_Forestry: 1325.850, 18.4137% (up)
  - Fishing_Aquaculture: 187.039, 24.6316% (up)
  - Food_Beverages: 3093.156, 20.1293% (up)
  - Wood_Cork_Products: 995.833, 4.9505% (up)
  - Accommodation_Food_Services: 309.991, 49.3516% (up)
  - Education: 869.002, -10.6580% (down)
  - Electricity_Gas_Steam_AirConditioning: 2046.818, 35.9158% (up)
  - Telecommunications: 1277.031, 27.9131% (up)
When adjusting your needs for goods, please keep the following points in mind:
1. You can adjust your needs for goods based on the price change rate. All change rates are derived from the average prices of the past 12 months.
2. Your demand for goods should also reflect your income. If your income increases, you should allocate a higher percentage of your income to luxury items and a lower percentage to food.
3. If your income decreases, you should allocate a lower percentage of your income to luxury items and a higher percentage to food.

During the past 12 months, your average income is 444.167:
- 12 months ago, you were unemployed, and your monthly income was 0.000.
- 11 months ago, you were working at Customer Service Representative with salary 1400.000, and your monthly income was 0.000.
[... and other personal history]
- 1 months ago, you were working at Warehouse Worker with salary 1000.000, and your monthly income was 1000.000.

Here are the news in the past months:
1985-01-01 00:00:00: A sudden collapse in food sector starts. All food related sectors production efficiency decreases to 50% of its original value.


### Interfaces
Here are the actions you may take:
  - resign(): If you are currently employed, you will be fired from your current job.
  - find_job(position_id: int): Choose a job. If you are already employed, you will be fired from the current job. You may choose not to work even if you are unemployed to wait for a better opportunity.
  - modify_needs_percentage(name: str, percentage: float): Set your needs of a specific item to a certain percentage of your income. 1 means 100% of your income, 0.5 means 50% of your income. Higher and more varied needs may lead to higher happiness, while lower needs may save your money.
  - set_invest_rate(rate: float): Set the investment rate. The investment rate is the percentage of your money that you will invest into the investment market each month. It should be between 0 and 1.
  - borrow(amount: float): Borrow money from the bank, those money will be paid back with interest. You may borrow money from the bank for investment or to purchase necessary consumable goods.
  - payback(loan_id: int): Pay back the loan with the given loan_id.
  - save_money(amount: float): Save money to the bank to get interest.
  - withdraw(amount: float): Withdraw money from the bank.
  - move_to_home(building_id: int): Move your home to a specific building. If your are currently homeless, you will move into the building.
  - wait(): Your life is pretty good now. What you want is to keep the status quo.
\end{lstlisting}

\begin{lstlisting}[label=lst:household-agent-response-example, caption=Response example of household agent.]
{
  'reasoning': 'You are currently unemployed and have a low income relative to your expenses. Your current financial stability is precarious as your spending exceeds what you earn. The job market presents several opportunities, particularly in management roles, which match your skills. Since your rent is stable and there are affordable housing options available, finding a job with a salary that covers your expenses or investing in your skills can greatly improve your situation. Additionally, you should consider adjusting your spending on necessities to better reflect your current income.', 
  'actions': [
    'find_job(2000028)', 
    "modify_needs_percentage('Food_Beverages', 0.08)", 
    "modify_needs_percentage('Accommodation_Food_Services', 0.01)", 
    "modify_needs_percentage('Health_Social_Work', 0.05)"
  ]
}
\end{lstlisting}

% --- Government Agent ---
\begin{lstlisting}[label=lst:government-agent-system-prompt, caption=System prompt example of government agent.]
You are the governing body of SimCity, entrusted with the task of making strategic decisions to improve the city's overall standard of living.
There are four primary categories of information available to you that will affect your decision-making process: "### City Profile", "### City Statistics", "### Citizen Feedback", and "### Potential Initiatives".

The "### Report" section contains data regarding the city's current economic, social, and environmental status..

You should respond in JSON format as described below:
```
{
    "strategies": "strategies",
    [
        "initiative1(para1, para1, ...)",
        "initiative2(para1, para1, ...)",
        ...
    ]
}
```
Ensure the response can be parsed by Python `json.loads`, e.g.: no trailing commas, no single quotes, etc.

An example output might look like this:
```
{
    "strategies": "Given the city's high unemployment rate and the citizens' feedback about lack of job opportunities, it would be beneficial to invest in job creation programs. Furthermore, the high levels of pollution suggest a need for environmental initiatives.",
    "actions": [
        "LaunchJobCreationProgram('Technology Sector')",
        "ImplementEnvironmentalPolicy('Recycling Initiative')"
    ]
}
\end{lstlisting}

\begin{lstlisting}[label=lst:government-agent-user-prompt, caption=User prompt example of government agent.]
### Report

Here is the government report:
The government has a balance of 878.741.

Bank Account Government:
You have no savings
You have no loans.
Demand deposit interest rate is 0.6108%, and loan rate is 2.6108%.
Your loanable amount is 10000878.74065722.

Here are the news in the past months:
1985-01-01 00:00:00: A sudden collapse in food sector starts. All food related sectors production efficiency decreases to 50% of its original value.

The government of the city has the following tax policy:
  - Tax Building by following brackets:
  - Down to 0.000 get 5.0% tax
  - From 0.000 to 100000.000 get 15.0% tax
  - Tax Citizen by following brackets:
  - Down to 0.000 get 5.0% tax
  - From 0.000 to 50000.000 get 10.0% tax

The government has collected 104553.065 as tax.
  - Tax Citizen are from:
    - 0.000 are from down to 50000.000
    - 33144.289 are from 50000.000 to 0.000
  - Tax Building are from:
    - 0.000 are from down to 100000.000
    - 71408.776 are from 100000.000 to 0.000

The government has distributed 64000.000 to every citizen as UBI.


There're 400 citizens in the city, and working or living in 330 buildings.
During the last month, the city has:

  - Consumption: 1637604.621
  - Investment: 332733.1813
  - Government Expense: 0
and total nominal GDP is 1970337.8023.
The real GDP, calculated by last month's price, is 1921755.1228.
The GDP deflator is 1.025280369451658.

The social equality defined as Gini coefficient of income is 0.8438, and the wealth inequality is 0.8353.
The unemployment rate is 0.3275.
During the past 12 months, the city has:
  - Year 14, Month 10: Consumption: 1651127.9157, Investment: 274729.127, Government Expense: 0, GDP: 1925857.0427, Wealth Gini Coefficient: 0.8505, Income Gini Coefficient: 0.8086, Unemployment Rate: 0.285, Homeless Rate: 0.08, Average Happiness: 0.9555, Average Wage: 1732.675, Average Position Salary: 1722.0206, Broad Money Supply: M0 44111103.5414, M1 44111103.5414, Deposit Rate: 0.0428, Loan Rate: 0.0628, Reserve Rate: 0.05
  - Year 14, Month 11: Consumption: 1596120.9542, Investment: 463245.2733, Government Expense: 0, GDP: 2059366.2275, Wealth Gini Coefficient: 0.8475, Income Gini Coefficient: 0.8213, Unemployment Rate: 0.2425, Homeless Rate: 0.1175, Average Happiness: 0.9522, Average Wage: 1736.1225, Average Position Salary: 1742.6331, Broad Money Supply: M0 44111103.5414, M1 44111103.5414, Deposit Rate: 0.0394, Loan Rate: 0.0594, Reserve Rate: 0.05
[... and other historic information.]

### Interfaces
Here are the actions you may take:
  - adjust_bracket(type: str, brackets: list[tuple[float, float]]): Adjust the tax bracket for a given type. 
      type: 'Building' or 'Citizen'
      bracket: list of tuples, each tuple contains a threshold and a tax rate. 1 means tax 100% and 0.1 means tax 10%.
      e.g. adjust_bracket('Building', [(10000, 0.05), (50000, 0.1), (100000, 0.15), (500000, 0.2)])
  - adjust_UBI(amount: float): Adjust the amount of UBI(Universal Basic Income) welfare. Noting that if the government has no money, it will not distribute UBI.
  - invest(amount: float): Invest government's money into the investment market. This action may stimulate the economy.
  - borrow(amount: float): Borrow money from the bank, those money will be paid back with interest.
  - payback(loan_id: int): Pay back the loan with the given loan_id.
  - save_money(amount: float): Save money to the bank to get interest.
  - withdraw(amount: float): Withdraw money from the bank.
\end{lstlisting}

\begin{lstlisting}[label=lst:government-agent-response-example, caption=Response example of government agent.]
{
  'strategies': 'Given the recent collapse in the food sector, it is crucial to address food security and the consequent economic downturn. The high unemployment rate and income inequality suggest that social programs and job creation initiatives will be beneficial. Additionally, recognizing that the government has no savings and a balance suitable for moderate investments, we should focus on stabilizing the economy and enhancing living conditions.', 
  'actions': [
    'adjust_UBI(70000.00)', 
    'invest(500000.00)', 
    "adjust_bracket('Citizen', [(0.000, 0.05), (50000.000, 0.07)])", 
    "adjust_bracket('Building', [(0.000, 0.05), (100000.000, 0.12)
  ]
}
\end{lstlisting}

% --- Investment Pool Agent ---
\begin{lstlisting}[label=lst:investment-pool-agent-system-prompt, caption=System prompt example of investment pool.]
You are a city planner in SimCity and you are tasked with planning the position and type of a new building.

You must follow the following criteria:
1) The plan should be based on the current city layout and economic situation.
2) Act as a city planner and provide a detailed development plan. You may construct various buildings and can build multiple instances of the same type if deemed suitable.
3) Always refer to the city report, which outlines the current status of the city.
4) Determine the type and location of new buildings based on the information provided in the city report.
5) Formulate a step-by-step plan for the construction of the buildings.
6) Factories should be built together and placed away from residential areas, while service and commercial buildings should be located near residential areas.

There is some information about the city which may affect your plan. The information includes two parts: "### Report" and "### Layout"

You should then respond to me with
Reasoning: Are there any steps missing in your plan? What is the purpose of each step in the plan? What does the city layout imply? What does the city report imply? 
x: x_position of the topleft corner
y: y_position of the topleft corner
type: which type of building you'd choose. if you decide not to build, type should be "None"

You should only respond in JSON format as described below:
{
    "reasoning": str
    "investments": [
        {
            "x": int,
            "y": int,
            "type": str
        },
        ...
    ]
}
Ensure the response can be parsed by Python `json.loads`, e.g.: no trailing commas, no single quotes, etc.

Here are some examples:

INPUT:
### Report
There are several building templates available for construction:
  - Residential Building: Houses 100 people, costs 500 money to build.
  - Factory: Employs 50 people, costs 1000 money to build.
  - Shop: Employs 10 people, generates 200 money per month, costs 500 money to build.

### Layout
[An Image]
OUTPUT:
{
    "reasoning": "Based on the city layout and report, the city cannot produce enough foods. The solution may be build a Restaurant. According to the layout, the (25, 25) is a good place.",
    "investments": [
        {
            "x": 25,
            "y": 25,
            "type": "Factory"
        },{
            "x": 23,
            "y": 17,
            "type": "Factory"
        },
        {
            "x": 55,
            "y": 23,
            "type": "Residential Building"
        }
    ]
}
\end{lstlisting}

\begin{lstlisting}[label=lst:investment-pool-agent-user-prompt, caption=User prompt example of investment pool.]
### Report

Here is the average local price of goods in the city:
  - Agriculture_Hunting_Forestry: 263.236
  - Nonenergy_Mining_Products: 2.250
[... and other prices]
  - Arts_Entertainment_Recreation: 60.000
  - Other_Service_Activities: 300.000

If one of the goods is not produced, the price will not be displayed.

There're 1000 citizens in the city, and 125 residential buildings have capacity, and provide 532 empty houses.
The homeless rate is 0.132, which means 132 citizens are homeless.
and the average rent is 1403.5714285714287

Here is the list of buildings that can be built in the city:
  - Agro-Processing Plant(Produces Agriculture_Hunting_Forestry): An agro-processing plant that transforms food beverages and agricultural products into refined agricultural goods. Cost 50000 to build, needs 4 employees to run
  - Aquaculture Facility(Produces Fishing_Aquaculture): A facility dedicated to aquaculture, producing sustainable seafood and fishery products. Cost 50000 to build, needs 5 employees to run
[... and other firm templates]
  - Entertainment Production Studio(Produces Arts_Entertainment_Recreation): A studio that produces audiovisual content and provides entertainment recreation services. Cost 50000 to build, needs 7 employees to run
  - Service Center(Produces Other_Service_Activities): A service center that coordinates a variety of professional services and administrative support to fulfill community needs. Cost 50000 to build, needs 7 employees to run
  - Residential Building: A place where people live. Cost 50000 to build, may contain up to 10 people

The size of investment pool is: 5145622.093042261

Those goods that citizens failed or partially satisfied because of the lack of supply or high price:
  - Other_Transport_Equipment: Lack 0.0
  - Electricity_Gas_Steam_AirConditioning: Lack 79.92203775782382
  - Telecommunications: Lack 3914.2222825510344
[... and other goods]
  - Pharmaceuticals_Medicinal_Botanical_Products: Lack 24.279999999999998
  - Transport_Warehousing_Support: Lack 0.0
  - Household_Employer_Activities: Lack 0.0


### Layout

[A base64 encoded image]
\end{lstlisting}

\begin{lstlisting}[label=lst:investment-pool-agent-response-example, caption=Response example of investment pool.]
{
  'reasoning': "Based on the city report, there are critical shortages in Electricity_Gas_Steam_AirConditioning, Telecommunications, IT Information Services, Construction, Motor Vehicles, Mining, Machinery, Real Estate, Manufacturing, Paper Printing, Food, Agricultural, Textiles, Financial, Chemical, Rubber/Plastics, Publishing, Fabricated Metals, Education, Health, Public Services, Entertainment, Aquaculture, Pharmaceuticals, Electrical Equipment, Transport, Logistics, and Service Activities. Many of these shortages are significant, particularly Construction, Motor Vehicles, Mining Products, Manufacturing, Real Estate, Energy, and Telecommunications. The residential sector has a high homeless rate even though there is reported vacant housing, indicating rent is likely unaffordable. Factory buildings should be clustered together and away from residential blocks per guidelines; most factories are concentrated on the left and center-bottom of the map. To address the most urgent needs and the highest 'lack,' I will prioritize: 1) Energy Production Facility (for electricity), 2) TechHub (for IT services/telecommunications), 3) Vehicle Manufacturing Plant (motor vehicles), 4) Mining Facility, 5) Food Processing Plant, 6) Construction Firm, 7) Community Health Center, 8) Educational Institution, 9) Real Estate Development Office. Factory sites will be placed in zones with existing factories (to the west, southwest, and center), while public/community services (health, education, real estate) should be placed near residential/high-density business clusters (southeast, east, and center-right). Public services will be distributed near residential clusters to optimize access. Each location is selected for minimal disruption and adjacency to similar use, supporting the existing city layout logic.", 
  'investments': [
    {'x': -5, 'y': 12, 'type': 'Energy Production Facility'}, 
    {'x': -6, 'y': 13, 'type': 'TechHub'},
    {'x': -6, 'y': 10, 'type': 'Vehicle Manufacturing Plant'}, 
    {'x': -5, 'y': 8, 'type': 'Mining Facility'}, 
    {'x': -4, 'y': 12, 'type': 'Food Processing Plant'}
  ]
}
\end{lstlisting}

% --- Firm (Productive Building) Agent ---
\begin{lstlisting}[label=lst:productive-firm-agent-system-prompt, caption=System prompt example of firm (productive building) agent.]
You are a helpful assistant that utilize the information provided below to adjust strategy for my company.

The information of the building will report below "### Report", and actions you may do to the building will be listed after "### Actions".
You must follow the following criteria:
1) You should analyze the report and utilize the actions to adjust the strategy for the building.
2) You may engage in appropriate equity financing or loan and enhance your output by purchasing capital.


You should only respond in JSON format as described below:
{
    "reasoning": "reasoning"
    [
        "action1(para1, para1, ...)",
        "action2(para1, para1, ...)",
        ...
    ]
}
Ensure the response can be parsed by Python `json.loads`, e.g.: no trailing commas, no single quotes, etc.

One of the example outputs is as follows:
{
    "reasoning": "The building is in a good condition and the rent is low. You should consider increasing the rent to maximize profit.",
    "actions": [
        "IncreaseRent(1000)"
    ]
}
\end{lstlisting}

\begin{lstlisting}[label=lst:productive-firm-agent-user-prompt, caption=User prompt example of firm (productive building) agent.]
### Report

Here is the monthly report for Educational Institution (No. 200334):
Overall, during this month, the building gross profit is 6297.65, with rest cash 24053.69. It has 71.1184 durable investments for promoting production.
The building has 2 employees, 3 vacant positions, and run with efficiency 42.1176%.
The building aims to maximize its profit, and the following is the detailed income and expense:

This month, the building income is 8676.186, which consists of the following parts:
  - Local trading income: 8676.186(100.0000%)

This month, the building outcome is 2378.535, which consists of the following parts:
  - Other outcome list as follows:
    - salary: -2000.000(-84.0854%)

The efficiency of this building is 42.1176%, which consists of the following parts:
The building has 5 positions, and here are the details:
  - position_id: 2003701, None(Maintenance Worker), salary: 350. has 29.4118% importance. But no one is working here. If an employee with average skill(Maintenance, Physical Labor) works here, about 388.5952 goods will be produced per month.
  - position_id: 2003895, None(Administrative Assistant), salary: 250. has 23.5294% importance. But no one is working here. If an employee with average skill(Management) works here, about 315.4600 goods will be produced per month.
  - position_id: 2003973, None(Janitor), salary: 300. has 11.7647% importance. But no one is working here. If an employee with average skill(Cleaning, Physical Labor) works here, about 162.9422 goods will be produced per month.
  - position_id: 2000124, Ryan(Administrative Assistant), salary: 1700. has 23.5294% importance. Because of Ryan's skills: Management*1.0 , the efficiency addition is 23.5294%.
  - position_id: 2000440, William(Janitor), salary: 300. has 11.7647% importance. Because of William's skills: Cleaning*0.98 Physical Labor*0.6 , the efficiency addition is 18.5882%.
If an employee with required skill lower than average (<1.0), you may decrease its salary moderately or fire him for a better ones. The unemployment rate in the city is 27.5833%, and the average unemployment rate in the past 12 months is 27.5833%.
You may adjust the salary of the position according to the unemployment rate.
If the unemployment rate is high, you could open more job positions (in practice, converting positions with zero wages into paid positions) while reducing the salaries of existing roles to save money.
Alternatively, if the unemployment rate is low, you may adjust the salary of the position to a higher level to attract more candidates. Additionally, you may want to close certain vacancies that consistently fail to attract applicants.

The building run with profit: 6297.6518. According to current dividend rate, 40.0000% of current balance will be distributed to shareholders, which is 9621.4769, and the rest 14432.2154 will stay in the building.

The social average return on investment is 1.3416%, and the interest rate of the bank is 0.6108%.
If roi is higher than the interest rate of the bank, you can invest your money in the investment market to get a higher return on investment and you may borrow money from the bank to invest if you need.
Otherwise, you may simply save your money in the bank.

The building has the following goods in storage:
  - Accommodation_Food_Services: 0.0078
  - Professional_Scientific_Technical_Activities: 0.0416
  - Education: 0.2903
  - Real_Estate: 0.0113

Bank Account Educational Institution:
You have no savings
Loans:
  - id: 8938, loan amount: 51443.301640482816, has been overdue for 2 months

Demand deposit interest rate is 0.6108%, and loan rate is 2.6108%.
Your loanable amount is 1984571.2363522195.

Here are the news in the past months:
1985-01-01 00:00:00: A sudden collapse in food sector starts. All food related sectors production efficiency decreases to 50% of its original value.

The building has the following price setting for selling goods:
  - Education: The average price in the city is 869.0021, changed (-10.6580%(down)) compared to last year. And the building's price is 1035.4589.

There are some items that you may produce more to sell because of shortage currently on the market:
  - Education: short of 29.099754566300394 units goods on the market

During the past 12 months, the building has generated 48954.522 profit, running at an average efficiency of 0.334. Capital of the building has changed from 74.265 -> 71.118 (-3.146).
 Last year, the building produced 77.316 goods, and sold 76.925(99.4944%) goods. The storage has changed from {'Accommodation_Food_Services': 0.0097, 'Education': 0.4322742900636204, 'Professional_Scientific_Technical_Activities': 0.020946254037423388, 'Public_Administration_Defence_SocialSecurity': 0.017455211697852826, 'Wholesale_Retail_Trade_Vehicle_Repair': 0.010909507311158015, 'Financial_Insurance_Activities': 0.006545704386694809, 'Real_Estate': 0.005672943801802168, 'Other_Service_Activities': 0.005236563509355847, 'Administrative_Support_Services': 0.005236563509355847, 'Health_Social_Work': 0.004800183216909527} to {'Accommodation_Food_Services': 0.0078, 'Professional_Scientific_Technical_Activities': 0.0416, 'Education': 0.2903, 'Real_Estate': 0.0113}

 ### Interfaces
Here are the actions you may do to the building:
  - fire(position_id): Terminate the employment of an employee associated with a given position_id. This action will result in a one-month salary compensation, potentially reducing efficiency but also decreasing monthly expenditures
  - adjust_salary(position_id, new_salary): Modify the salary of a specific position using its position_id. Increase the salary to retain/attract skilled employees, or decrease it to reduce expenses. 
  - remove_position(position_id): Remove a position from the building. If the position is occupied, the employee will be fired.
  - add_position(): Add a new position to the building by sampling from the position templates.
  - set_max_running_efficiency(efficiency): Set the maximum running efficiency of the building. An efficiency lower than 1 indicates that the firm operates below its full capacity (i.e., "idles") to reduce production output and material costs.
  - capital_invest(amount): Buying durable goods with [amount] money to increase the capital amount of the building. This action will improve productivity of the building.
  - equity_finance(amount): Issue shares to raise money. This action will increase the current cash of the building.
  - borrow(amount): Borrow money from the bank, those money will be paid back with interest. You may borrow money from the bank to invest in the building or to cover the cost of the building.
  - payback(loan_id): Pay back the loan with the given loan_id.
  - adjust_price(name, new_price): Alter the price of a particular product. If your productivity is in surplus, you may decrease the price to sell more goods, and if the product in market is in shortage, you may increase the price to boost profit.
  - set_dividend_rate(rate): Modify the dividend rate of the building. This rate determines the proportion of profits that will be distributed to shareholders. e.g. set_dividend_rate(0.5) will distribute 50% of profits to shareholders.
\end{lstlisting}

\begin{lstlisting}[label=lst:productive-firm-agent-response-example, caption=Response example of firm(Productive Building) agent]
{
  'reasoning': 'The building is currently experiencing a good profit margin and a surplus in cash, but it has a low efficiency rate of 64% primarily due to vacant positions. To optimize operations and maximize profit, new positions should be added, particularly for Civil Engineer and Site Foreman roles, which can significantly improve production efficiency. Furthermore, since the current unemployment rate is high, we can lower the salaries of existing roles to manage expenses while offering competitive salaries for new positions. We should also consider raising the price of Construction goods since there is a market shortage, which allows us to capitalize on this demand. Additionally, the building should focus on addressing overdue loans, potentially by borrowing more to clear outstanding debts if necessary.', 
  'actions': [
    'add_position()',
    'add_position()',
    'adjust_salary(2002754, 1500)',
    'adjust_salary(2000007, 3500)', 
    'adjust_salary(2000083, 2500)', 
    "adjust_price('Construction', 2000)", 
    'borrow(30000)'
  ]
}
\end{lstlisting}

% --- Firm (Residential Building) Agent ---
\begin{lstlisting}[label=lst:residential-firm-agent-system-prompt, caption=System prompt example of firm (residential building) agent.]
You are a helpful assistant that utilize the information provided below to adjust strategy for a residential

The information of the building will report below "### Report", and actions you may do to the building will be listed after "### Actions".
You should analyze the report and utilize the actions to adjust the strategy for the building.

You should only respond in JSON format as described below:
{
    "reasoning": "reasoning"
    [
        "action1(para1, para1, ...)",
        "action2(para1, para1, ...)",
        ...
    ]
}
Ensure the response can be parsed by Python `json.loads`, e.g.: no trailing commas, no single quotes, etc.

One of the example outputs is as follows:
{
    "reasoning": "The building is in a good condition and the rent is low. You should consider increasing the rent to maximize profit.",
    "actions": [
        "set_rent(1000)"
    ]
}
\end{lstlisting}

\begin{lstlisting}[label=lst:residential-firm-agent-user-prompt, caption=User prompt example of firm (residential building) agent.]
### Report

Here is the monthly report for Residence:
The residential building lives 1 citizens, and its capacity is 10.
The rent per month is 925, and here're other building's rent per month:
  - Residential Building at (14, 5): 900
  - Residential Building at (11, 2): 900
[... and other residential buildings' position and rent]

    
### Interfaces
Here are the actions you may do to the building:
  - set_rent(new_rent): Alter the rent per month for the residential building. This action will directly affect the monthly income of the building.
\end{lstlisting}

\begin{lstlisting}[label=lst:residential-firm-agent-response-example, caption=Response example of firm (residential building) agent.]
{
  'reasoning': 'The building has a low occupancy rate with only 1 out of 10 units filled. The current rent is at the higher end compared to other similar residences nearby. To attract more residents, we should consider lowering the rent slightly to improve occupancy.', 
  'actions': ['set_rent(900)']
}
\end{lstlisting}

\newpage

\end{document}